\documentclass[11pt,preprint,showpacs]{revtex4}
\usepackage{graphicx}
\usepackage{dcolumn}
\usepackage{bm}
\usepackage{amsmath}
\usepackage{amssymb}
\usepackage{amsfonts}

\begin{document}

\title{A novel treatment of the proton-proton  Coulomb force in
elastic  proton-deuteron  Faddeev calculations}

\author{H.~Wita{\l}a}
\affiliation{M. Smoluchowski Institute of Physics, Jagiellonian
University,
                    PL-30059 Krak\'ow, Poland}

\author{R.~Skibi\'nski}
\affiliation{M. Smoluchowski Institute of Physics, Jagiellonian
University,
                    PL-30059 Krak\'ow, Poland}

\author{J.~Golak}
\affiliation{M. Smoluchowski Institute of Physics, Jagiellonian
University,
                    PL-30059 Krak\'ow, Poland}

\author{W.\ Gl\"ockle}
\affiliation{Institut f\"ur theoretische Physik II,
Ruhr-Universit\"at Bochum, D-44780 Bochum, Germany}

\date{\today}

\begin{abstract}
We propose a novel approach to incorporate the proton-proton (pp)
Coulomb force into the three-nucleon (3N) Faddeev calculations.
The main new ingredient is a 3-dimensional screened pp  Coulomb
t-matrix obtained by a numerical solution of the 3-dimensional
Lippmann-Schwinger (LS) equation. We demonstrate numerically and
provide analytical insight  that the elastic proton-deuteron (pd)
observables can be determined directly from the resulting on shell
3N amplitude increasing the screening radius. The screening limit
exists without the need of renormalisation not only for observables but 
for the elastic pd amplitude itself.
\end{abstract}


\maketitle \setcounter{page}{1}

\section{Introduction}
 \label{intro}

The inclusion of the Coulomb force into the analysis  of nuclear
reactions with more than 2 nucleons is a long standing problem.
The main reason is the  long-range nature of the Coulomb force which
prevents the application of the standard techniques developed for
short-range interactions. One possible way to avoid the
difficulties  including the Coulomb force is to use a screened
Coulomb interaction and to reach the pure Coulomb limit through
application of a renormalisation procedure~\cite{Alt78,Alt96,Alt94,Alt2002}.

The problem is especially timely to be solved  when considering the
interaction
of protons with deuterons below the pion production threshold. For
this 3N system using the Faddeev scheme high-precision numerical
predictions for different observables in elastic proton-deuteron
(pd) scattering and for the deuteron breakup reaction are being
obtained \cite{physrep96}, however, only under the restriction to
short-ranged nuclear interactions. The high quality of the
available pd data for both processes requires, however, in the theoretical
analysis the
inclusion of the Coulomb force into the calculations. Furthermore
the seminal progress \cite{epel2006} in the development of nuclear forces in
chiral effective field theory calls also for a very precise
solution of the pd scattering equations to test unambiguously these new
dynamical ingredients. This test can only be completely satisfactory if the
pp Coulomb force is perfectly under control.

 For the elastic pd
scattering first calculations, with modern nuclear forces and the
Coulomb force included, have been achieved in a variational
hyperspherical harmonic approach~\cite{kievski}. Only recently the
inclusion of the Coulomb force became possible also for the pd
breakup reaction~\cite{delt2005br}. In~\cite{delt2005br}, contrary
to~\cite{kievski} where the exact Coulomb force in coordinate
representation has been used directly, a screened pp Coulomb force
has been applied in momentum space and in a partial wave basis. In
order to get the final predictions which can be compared to the
data, the limit to the unscreened situation has been performed
numerically applying a renormalization to the resulting
on-shell amplitudes~\cite{delt2005el,delt2005br}. This allowed for the
first time to analyze high-precision pd breakup data and provided
a significant improvement of data description in cases where the
Coulomb force plays an important role~\cite{stephan}.

However, in spite of that substantial progress some important
questions remained unanswered. One concerns directly the results
of these calculations for two kinematically complete breakup
geometries: the pp quasi-free-scattering (QFS) configuration, in
which the not detected neutron is at rest in the laboratory system,
and the space-star (SST) geometry, in which all 3 outgoing
nucleons have the same  momenta (magnitudes)  in the plane which in the 3N
c.m. system is
perpendicular  to the incoming nucleon
momentum. The theoretical predictions based on nuclear forces only
show, that the cross sections for QFS and SST are quite stable
against changes of the underlying interactions, including also
three-nucleon forces~\cite{physrep96}. At energies below $\approx 20$~MeV
theory
underestimates the SST pd cross sections by $\approx 10\%$,
and overestimates the pp QFS cross sections by $\approx 20\%$,
respectively~\cite{physrep96,delt2005br}. With increasing energy  the
theoretical cross sections
come close to the data, which  indicates that the pp Coulomb
force is very probably responsible for these low energy
discrepancies. However, the Coulomb force effects found
in~\cite{delt2005br} are practically negligible for the pd SST
configuration and only slightly improve the description of the pp
QFS data~\cite{exp1,exp2,exp3,exp4,exp5}.

This inability to understand the pp QFS and pd SST cross sections
calls for reconsidering the  inclusion of the Coulomb force
into momentum space Faddeev calculations. One main concern in such
type of calculations is the application of a partial wave decomposition
to the long-ranged  Coulomb force. Even when screening is applied
it  seems reasonable to treat from the beginning the screened pp
Coulomb t-matrix without partial wave decomposition because the required
limit
 of vanishing screening leads necessarily to  a drastic increase
of the number of partial wave states involved.
 As an example we provide numbers for the exponential screening 
 $e^{-(\frac {r} {R})^n}$. Taking the screening radius $R=20$~fm and 
$n=4$ requires all $l \le l_{max}=10$ partial wave states to reproduce the 
$3$-dimensional pp screened Coulomb t-matrix at $E_p^{lab}=13$~MeV. 
Increasing the screening radius to $R=120$~fm requires  $l_{max} \approx 50$ which is 
 a big numerical challenge. Even more that would lead to an explosion of the 
number of 3N partial waves required for convergence.

Another problem concerns the treatment of the pp Coulomb
interaction in its proper coordinate, which is the relative
proton-proton distance, throughout the  calculations. Any
deviation from this restriction can cause effects which are
difficult to estimate.

In the present paper we propose a novel approach taking both concerns  
 into account to incorporate
the pp Coulomb force into the momentum space Faddeev calculations,
in which we apply directly the 3-dimensional screened pp Coulomb
t-matrix without relying on a partial wave decomposition.
3-dimensional solutions of the LS equation for different screening
radii are used to approach the final predictions. We demonstrate,
that even the physical elastic pd scattering amplitude has a well
defined screening limit and does not require renormalisation. In
contrast in case of pp scattering the scattering amplitude
requires renormalisation in the screening limit which,  however,
 has not to be applied when only observables have to be
addressed \cite{Witala08}.

In section \ref{form} we present details of 
the formalism and in section \ref{ampli} 
  the physical amplitude for elastic pd scattering. The
screening limit is discussed in section \ref{limit} and the results shown
in section \ref{results}. The summary is given in section \ref{sumary}. 
In various
appendices detailed expressions for different kernel elements
appearing in the Faddeev equation with the screened Coulomb force
are included and  different terms contributing to the on-shell
elastic scattering amplitude are given.

\section{Faddeev equations with screened pp Coulomb force}
\label{form}

The 3N pd system can be regarded in a particle basis or in the isospin
basis.
 The corresponding completeness relations are
\begin{eqnarray}
| pnp> < pnp| + | ppn> < ppn| + | npp> < npp| = 1\label{1} ~,
\end{eqnarray}
\begin{eqnarray}
\sum_{tT} | ( t 1/2 )T -1/2> < ( t 1/2) T -1/2| = 1 ~.
\label{2}
\end{eqnarray}
  In (\ref{2}) t is the 2-body isospin which together with the
 isospin $ \frac{1}{2}$ of the third particle is coupled to the total
isospin  T.
 We use the convention that
 the proton (neutron) has the isospin magnetic quantum number
$ - \frac{1}{2}  (\frac{1}{2})$. Since NN forces are usually given in an
 isospin basis we stick to that formulation also in the 3N system.
 Then a general NN force acts in the 3N system as
 \begin{eqnarray}
V = \sum_{t,T} \sum_{t',T'} | ( t 1/2 )T M_T> < ( t 1/2) T M_T| V | ( t'
1/2) T' M_T>
 < ( t' 1/2 ) T' M_T|\label{3} ~.
\end{eqnarray}
Here we assume that t is conserved, which in the pd system leads to
\cite{wit91}
\begin{eqnarray}
V& =&  |(01/2) 1/2 -1/2> V_{np}^{00} < ( 01/2) 1/2 -1/2|\nonumber\\
& + &  |(11/2) 1/2 -1/2> ( 1/3 V_{np}^{10} + 2/3 V_{pp}^{1-1})
 < ( 11/2) 1/2 -1/2|\nonumber\\
& + & |(11/2) 1/2 -1/2>  \frac{\sqrt{2}}{3}( V_{np}^{10} -V_{pp}^{1-1})
 < ( 11/2) 3/2 -1/2|\nonumber\\
& + & |(11/2) 3/2 -1/2> ( 2/3 V_{np}^{10} + 1/3 V_{pp}^{1-1})
 < ( 11/2) 3/2 -1/2|\nonumber\\
& + & |(11/2) 3/2 -1/2>  \frac{\sqrt{2}}{3}( V_{np}^{10} -V_{pp}^{1-1})
 < ( 11/2) 1/2 -1/2| ~.
\label{4}
\end{eqnarray}
 We defined in the usual manner
\begin{eqnarray}
V_{np}^{t0} & = &  < t 0| V | t0>\label{5} ~,\\
V_{pp}^{1 -1}  & = &  < 1 -1 | V | 1 -1>\label{6} ~.
\end{eqnarray} The latter decomposes into the strong part and the
pure Coulomb part,
 which we assume to be screened and parametrised by some parameter $R$
\begin{eqnarray}
V_{pp}^{1 -1} = V_{pp}^{strong} +V_{pp}^{c R}\label{7} ~.
\end{eqnarray}
 Eq.(\ref{4})  exhibits that charge independence
breaking leads necessarily
 to a coupling of $ T=1/2 $ and $ T=3/2$ states and the  complete action
 of the Coulomb force also requires  the presence of both total isospin
states.
Exactly the same decomposition holds for the 2N t-operator $ \hat t $
\begin{eqnarray}
\hat t & =&  |(01/2) 1/2 -1/2> t_{np}^{00} < ( 01/2) 1/2 -1/2|\nonumber\\
& + &  |(11/2) 1/2 -1/2> ( 1/3 t_{np}^{10} + 2/3 t_{pp}^{1-1})
 < ( 11/2) 1/2 -1/2|\nonumber\\
& + & |(11/2) 1/2 -1/2>  \frac{\sqrt{2}}{3}( t_{np}^{10} -t_{pp}^{1-1})
 < ( 11/2) 3/2 -1/2|\nonumber\\
& + & |(11/2) 3/2 -1/2> ( 2/3 t_{np}^{10} + 1/3 t_{pp}^{1-1})
 < ( 11/2) 3/2 -1/2|\nonumber\\
& + & |(11/2) 3/2 -1/2>  \frac{\sqrt{2}}{3}( t_{np}^{10} -t_{pp}^{1-1})
 < ( 11/2) 1/2 -1/2| ~. \label{8}
\end{eqnarray}

It is not difficult to see~\cite{wit91} that this is consistent with the
2-body Lippmann
 Schwinger equation
\begin{eqnarray}
\hat t = V + V G_0 \hat t\label{9}
\end{eqnarray}
 when expanded into the 3N isospin basis.

It is now convenient to define the three 3N isospin states
\begin{eqnarray}
|\gamma_0> & = & |(01/2) 1/2 -1/2> \cr |\gamma_1> & = & |(11/2)
1/2 -1/2> \cr |\gamma_2> & = & |(11/2) 3/2 -1/2> ~. \label{10}
\end{eqnarray}
Then $\hat t $  appears as
\begin{eqnarray}
\hat t = \sum_{\gamma,\gamma'} | \gamma> t_{\gamma,\gamma'} <
\gamma'|\label{11}
\end{eqnarray}
with
\begin{eqnarray}
t_{\gamma,\gamma'} = \delta_{ tt'} t_t^{TT'}\label{12}
\end{eqnarray}
and $ t_t^{TT'}$ can be read of from  (\ref{8}).

We use the Faddeev equation in the form~\cite{physrep96}
\begin{eqnarray} T| \Phi> = \hat t P | \Phi> + \hat t P G_0 T|
\Phi>\label{13}
\end{eqnarray}
 where $ P $ is defined in terms of transposition operators,
 $ P =P_{12} P_{23} + P_{13} P_{23} $, $G_0$ is the free 3N propagator,
 $ |\Phi>$ the initial state composed of a deuteron state and a momentum
 eigenstate of the proton.
Knowing $T| \Phi>$ the breakup as well as the elastic pd scattering
amplitudes
 can be gained by quadratures in the standard manner~\cite{physrep96}.

Expanding $T| \Phi>$
\begin{eqnarray}
T| \Phi> &=& \sum_{\gamma} | \gamma> T_{\gamma} \label{14}
\end{eqnarray}
(\ref{13}) turns into
\begin{eqnarray}
T_{\gamma}  &=&  \sum_{ \gamma'} t_{\gamma \gamma'}
< \gamma'| P |\Phi>  +
  \sum_{ \gamma'} t_{\gamma \gamma'}< \gamma'| P G_0 |
\gamma''> T_{\gamma''}  ~. \label{15}
\end{eqnarray}
In a more detailed notation we use (\ref{12}) and define
\begin{eqnarray}
T_{\gamma} \equiv T_t^T\label{16}\\
<\gamma|P|\gamma'> = \delta_{ TT'} F( t t' T) P_{t t'} \equiv
\delta_{T T'} P_{t t'}^T\label{17}
\end{eqnarray}
where $ P_{t t'} $ acts only in spin-momentum space,
\begin{eqnarray}
P_{t t'} = P_{12} P_{23} + ( -)^{t + t'} P_{13}
P_{23}|_{spin-momentum}\label{18}
\end{eqnarray}
 and $ F( t t' T) $ is given by
\begin{eqnarray}
  F( t t' T) = ( -)^{t'} \sqrt{2 t+1} \sqrt{2 t' +1} \times
\left\{
\begin{matrix}
t  & 1/2  & 1/2 \cr
 t' & T    & 1/2 \cr \end{matrix}
\right\}\label{19} ~.
 \end{eqnarray}
  For the convenience of the reader the nonzero values are

\begin{eqnarray}
F(1 1 1/2 ) &= & - \frac {1} {2} \\
F(1 1 3/2 ) &=& 1\\
F(1 0 1/2 ) &=&  \frac{\sqrt{3}}{2}\\
F(0 0 1/2 ) &=& - \frac {1} {2} \\
F(0 1 1/2 ) &=& - \frac{\sqrt{3}}{2}
\end{eqnarray}

This leads to the more explicit form of the  coupled set of
equations (\ref{15})
\begin{eqnarray}
T_t^T =   t_{t}^{T \frac{1}{2}} P_{t0}^{\frac{1}{2}} | \phi>
 +   \sum_{T'} t_t^{TT'} \sum_{t''} P_{t t''}^{T'} G_0 T_{t''}^{T'} ~.
\label{25}
\end{eqnarray}
where  we used the isospin independence of the free propagator $G_0$
and \begin{eqnarray} |\Phi> = | \gamma_0> | \phi>\label{26} ~.
\end{eqnarray}
 This set (\ref{25}) would be the starting point for a
calculation using momentum vectors
instead of an angular momentum decomposition.
In view of forthcoming calculations we provide that set as follows
\begin{eqnarray}
T_0^{1/2} &= &  - \frac {1} {2} t_{np}^{00} P_{00} | \phi>  - \frac
{1} {2}
t_{np}^{00} P_{00} G_0 T_0^{1/2} - \frac{\sqrt{3}}{2} t_{np}^{00}
P_{01} G_0 T_1^{1/2} ~, 
\label{27}
\end{eqnarray}

\begin{eqnarray}
T_1^{1/2} &= &  \frac{\sqrt{3}}{2}( \frac{1}{3}  t_{np}^{10} +
\frac{2}{3} t_{pp}^{1-1} ) P_{10} | \phi>
  +   \frac{\sqrt{3}}{2}( \frac{1}{3}  t_{np}^{10} +
\frac{2}{3} t_{pp}^{1-1} ) P_{10} G_0 T_0^{1/2}\cr
 & - & \frac {1} {2}
(\frac{1}{3} t_{np}^{10} + \frac{2}{3} t_{pp}^{1-1} ) P_{11} G_0
T_1^{1/2} 
  +  \frac{\sqrt{2}}{3} ( t_{np}^{10} - t_{pp}^{1-1} )
P_{11} G_0 T_1^{3/2} ~, 
\label{28}
\end{eqnarray}

\begin{eqnarray}
T_1^{3/2} &= &  \frac{1}{\sqrt{6}}(  t_{np}^{10} - t_{pp}^{1-1} )
P_{10} | \phi> 
  +   \frac{1}{\sqrt{6}} ( t_{np}^{10} - t_{pp}^{1-1} ) P_{10} G_0
T_0^{1/2}\cr
 & - & \frac{1}{3 \sqrt{2}} (t_{np}^{10} - t_{pp}^{1-1} ) P_{11} G_0
 T_1^{1/2} 
 +  ( \frac {2} {3} t_{np}^{10} + \frac {1} {3}  t_{pp}^{1-1} ) P_{11} G_0
 T_1^{3/2} ~.
\label{29}
\end{eqnarray}
Note $t_{pp}^{1-1}$ is driven by the strong and the screened pp
Coulomb force (\ref{7}).

Here, however,  we apply the angular momentum decomposition and
define the basis states
\begin{eqnarray}
|pq a> \equiv |pq ( ls) j ( \lambda \frac{1}{2} ) I ( j I)
JM>\label{parw}
\end{eqnarray}
where $ p,q$ are the standard Jacobi momenta,  $ (ls)j $ refers to the
2-body subsystem,
 $ ( \lambda \frac{1}{2} ) I$ to the third particle and  $JM$
 is the total angular momentum and its magnetic quantum number.
 In that partial wave basis the set (\ref{25}) reads
\begin{eqnarray}
<pq a| T_t^T > & = &  \sum_{a'} \int dp' p'^ 2 dq' q'^ 2 <pq a| t_t^{T
\frac{1}{2}} | p'q'a'>
 < p'q'a' | P_{t0}^{\frac{1}{2}} | \phi>\cr
&  + &  \sum_{ T' t"} \sum_{a'} \int dp' p'^ 2 dq' q'^ 2 <pq a|
t_t^{T T'} | p'q'a'> < p'q'a' |P_{t{t"}}^{T'} G_0 T_{{t"}}^{T'} >
~. \label{parwset}
\end{eqnarray}
This form allows us to separate the lower partial waves in the 2-body
subsystem,
 where the strong force acts together with the Coulomb force, from the
higher partial waves,  where only the
 pure pp Coulomb t-matrix is present.

Thus for the higher partial waves, denoted as $ |pq a^{>} > $, we obtain
\begin{eqnarray}
<pq a^{>} | T_0^{\frac{1}{2}} > & = &  0\label{parw0}\\
<pq a^{>} | T_1^T > & = &  \sum_{a'}  \int dp' p'^ 2 dq' q'^ 2 <pq
a^{>} | t_{1c}^{T \frac{1}{2}} | | p'q'a'> < p'q'a' |
P_{10}^{\frac{1}{2}} | \phi>\cr &   + &  \sum_{ T' t"} \sum_{a'}
\int dp' p'^ 2 dq' q'^ 2 <pq a^{>} | t_{1c}^{T T'} |  p'q'a'> <
p'q'a' | P_{1t"}^{T'} G_0  | T_{t"}^{T'}> \label{parw1}
\end{eqnarray}
where $ t_{1c}^{T T'} $  up to a factor is given by the screened
pure
 Coulomb t-matrix $ t_{pp}^{cR} $ contained in $ t_{pp}^{1 -1} $. Thus
\begin{eqnarray}
t_{1c}^{T T'} = f^{ 1T 1T'} t_{pp}^{cR}\label{31}
\end{eqnarray}
and  $ f^{ 1T 1T'} $ can be read of from (\ref{8}) as $ f^{
1\frac{1 }{2} 1\frac{1 }{2}} = \frac{2}{3} = 2 f^{ 1\frac{3 }{2}
1\frac{3 }{2}}, f^{ 1\frac{1 }{2} 1\frac{3 }{2}} = f^{ 1\frac{3
}{2} 1\frac{1 }{2}} = - \frac{ \sqrt{2}}{3} $ .

Inserting  the factors $f^{ 1T 1T'}$ it  simply follows from
(\ref{parw1}) that
\begin{eqnarray}
 <pq a^{>} | T_1^{\frac{1}{2}} > + \sqrt{2} <pq a^{>} | T_1^{\frac{3}{2}}> =
0 ~. \label{32}
\end{eqnarray}
The physical meaning of that relation is trivially evident in the particle
picture.
 One has
\begin{eqnarray}
  | \gamma_1>  & =  & \frac{1}{\sqrt{6}}( | ppn> + | pnp> - 2 |
npp>)\label{33}\\
| \gamma_2>  & = &  \frac{1}{\sqrt{3}}( | ppn> + | pnp> +  |
npp>)\label{34}
\end{eqnarray}
where the sequence  $ pnp $ for instance corresponds to the
sequence of particle numbers $ 1,2,3$.

Inserting the definitions  (\ref{14}) and (\ref{16}) the relation
(\ref{32}) can be rewritten as
\begin{eqnarray}
<pq a^{>} | ( <\gamma_1| + \sqrt{2} < \gamma_2| ) T> = \sqrt{\frac{3}{2}}
<pq a^{>} |
 (  <ppn| + < pnp| ) |T> ~. \label{35}
\end{eqnarray}
Now the $T$-amplitude projected on high angular momenta is
proportional to the Coulomb $ pp $ t-matrix, which acts in the
23-subsystem. Consequently the overlap in (\ref{35}) is zero.

Now using (\ref{32}) and (\ref{31}) the relation (\ref{parw1}) for $ < pq
a^ >
| T_1^
{\frac{1}{2}} >$ simplifies to
\begin{eqnarray}
 < pq a^ > | T_1^ {\frac{1}{2}} > & = &  \frac{1}{\sqrt{3}} \sum_{a'} \int
dp' p'^ 2 dq' q'^ 2
 < pq a^ > | t_{pp}^ {cR} |p' q' a'> < p' q'  a'| P_{10} | \phi>\cr
& + &  \frac{1}{\sqrt{3}}\sum_{a'}  \int dp' p'^ 2 dq' q'^ 2 < pq a^ > |
t_{pp}^ {cR} |p' q' a'>
 <p' q'  a'| P_{10}  G_0 | T_0^ {\frac{1}{2}} >\cr
& - &  \frac{1}{3}\sum_{a'}  \int dp' p'^ 2 dq' q'^ 2 \sum_{a''} \int dp''
p''^ 2 dq'' q''^ 2
 < pq a^ > | t_{pp}^ {cR} | p' q' a'>\cr
& &  <p' q'  a'| P_{11} G_0 | p'' q'' a^ {''<} >
 < p'' q'' a^ {''<} ( | T_1^ {\frac{1}{2}}> + \sqrt{2} |T_1^{\frac{3}{2}}>)
~. \label{36}
\end{eqnarray}

Because of (\ref{32}) there are no contributions of the higher
partial wave components and only the lower ones, denoted as $ < p
q a^ <  | $, are applied onto the amplitudes $ |T>$.  Due to
(\ref{32}) the same is true for $< pq a^> | T_1^{\frac{3}{2}}>$.
For the sake of a simpler notation we introduce $ | A
> \equiv | pq a^ < > $  and $ | B> \equiv |p q a^{>}> $ in the following and
 drop the summation and
integration signs.

Now we turn to the lower partial waves. From (\ref{27}) - (\ref{29})  and
inserting $ t_t^ {TT'}$, $P_{tt'}^ T$ explicitly  we get
\begin{eqnarray}
<A |T_0^ { 1/2}> &=& - \frac {1} {2} <A  | t_{np}^{00}  P_{00}| \phi>\cr &- &
\frac {1} {2}   <A  | t_{np}^{00}  P_{00}G_0 |A'>  <A' | T_{0}^ { 1/2}> \cr &
- & \frac{\sqrt{3}}{2}  <A | t_{np}^{00} P_{01} G_0 | A'> < A' |
T_{1}^ { 1/2}>\cr & - &\frac{\sqrt{3}}{2}  < A | t_{np}^{00}
P_{01} G_0  | B'> < B'|  T_{1}^ { 1/2}> ~, 
\label{37} \\
 < A | T_{1}^ { 1/2}>  &=&  \frac{1}{2 \sqrt{3}} <A |( t_{np}^{10} + 2
t_{pp}^{1-1}) P_{10} | \phi>\cr & + & \frac{1}{2 \sqrt{3}}   <A |(
t_{np}^{10} + 2 t_{pp}^{1 -1})   P_{10} G_0  |A'>  <A' | T_{0}^ {
1/2}>\cr & - & \frac {1} {6}  < A |( t_{np}^{10} + 2 t_{pp}^{1 -1}) P_{11}
G_0  |A'> <A' | T_{1}^ { 1/2}> \cr & - & \frac {1} {6}  <A |( t_{np}^{10} +
2 t_{pp}^{1 -1}) P_{11} G_0  |B'> < B'| T_{1}^ { 1/2}> \cr & + &
\frac{\sqrt{2}}{3}  < A  | (t_{np}^{10} -t_{pp}^{1 -1}) P_{11}G_0
|A'> <A'| T_{1}^ { 3/2}> \cr & + & \frac{\sqrt{2}}{3} < A |
(t_{np}^{10} -t_{pp}^{1 -1}) P_{11}G_0
 |B'> <B' | T_{1}^ { 3/2}> ~, 
\label{38}\\
 <A | T_{1}^ { 3/2}>  &=& \frac{1}{\sqrt{6}} <A | (t_{np}^{10} -t_{pp}^{1
-1}) P_{10} | \phi>\cr
  & + &\frac{1}{\sqrt{6}}  <A  | (t_{np}^{10} -t_{pp}^{1 -1}) P_{10}G_0
|A'> < A' |T_{0}^ { 1/2}>\cr
   & - & \frac{ 1}{3 \sqrt{2}}  < A | (t_{np}^{10} -t_{pp}^{1 -1}) P_{11}G_0
|A' > < A' |T_{1}^ {1/2}>\cr
& + & \frac {1} {3}  < A | (2 t_{np}^{10} +t_{pp}^{1 -1}) P_{11}G_0 |A'> <A'
|T_{1}^ { 3/2}>\cr & - &  \frac{ 1}{3\sqrt{2}}  <A | (t_{np}^{10}
- t_{pp}^{1 -1}) P_{11}G_0 | B'> < B' |T_{1}^ { 1/2}>\cr
 & + &  < A | ( \frac{2}{3} t_{np}^{10} + \frac{1}{3} t_{pp}^{1 -1})
P_{11}G_0  | B'> < B'|T_{1}^ { 3/2}> ~. \label{39}
\end{eqnarray}
Note $t_{pp}^{1-1}$ is the pp t-matrix driven by the strong and
screened pp Coulomb force and is projected  on a certain set of
low partial waves.

In (\ref{38}) and (\ref{39}) one can use
(\ref{32}) to eliminate $ < B' |T_{1}^ { 3/2}> $ leading to
\begin{eqnarray}
< A  | T_{1}^ { 1/2}>  &=&  \frac{1}{2 \sqrt{3}} < A |( t_{np}^{10} + 2
t_{pp}^{1-1}) P_{10} | \phi>\cr
& + & \frac{1}{2 \sqrt{3}}   < A  |(  t_{np}^{10} + 2 t_{pp}^{1 -1}) P_{10}
G_0  | A'>  < A'
 | T_{0}^ { 1/2}>\cr
& - & \frac {1} {6}  < A |( t_{np}^{10} + 2 t_{pp}^{1 -1}) P_{11} G_0  |A'>
<A' | T_{1}^ { 1/2}> \cr & + & \frac{\sqrt{2}}{3} <A  |
(t_{np}^{10} -t_{pp}^{1 -1}) P_{11}G_0 | A'> <A' | T_{1}^ {
3/2}>\cr & - & \frac{1}{2} <A | t_{np}^{10}  P_{11}G_0 |B'> < B'| T_{1}^ {
1/2}> ~,
\label{40}\\
 <A  | T_{1}^ { 3/2}>  &=& \frac{1}{\sqrt{6}}
< A  | (t_{np}^{10} -t_{pp}^{1 -1}) P_{10} | \phi>\cr
& + &\frac{1}{\sqrt{6}}  < A  | (t_{np}^{10} -t_{pp}^{1 -1}) P_{10}G_0  |A'>
<A' |T_{0}^ { 1/2}>\cr
& - & \frac{ 1}{2\sqrt{2}}  <A | (t_{np}^{10} -t_{pp}^{1 -1}) P_{11}G_0  |A'
> < A' |T_{1}^ {1/2}>\cr
& + &  \frac {1} {3}  <A | (2 t_{np}^{10} +t_{pp}^{1 -1}) P_{11}G_0 |A'> <A'
|T_{1}^ { 3/2}>\cr & - &  \frac{ 1}{\sqrt{2}}  <A |  t_{np}^{10}
P_{11}G_0  | B'> <  B' |T_{1}^ { 1/2}> ~. \label{41}
\end{eqnarray}
The sum over the high partial waves acts on $ |T_{1}^ { 1/2}> $
which, according to (\ref{36}), is driven by the screened Coulomb
t-matrix $ t_{pp}^ {cR} $. In order to avoid approximations we sum
up the high partial waves to infinity and add and subtract the
projection of $ t_{pp}^{cR}$ on the finite number of low partial
waves. Thus we put
\begin{eqnarray}
&& | B'> < B' |T_{1}^ { 1/2}>  \equiv |\tilde T_{1}^ { 1/2}>
\label{42}
\end{eqnarray}
where according to (\ref{36}) \begin{eqnarray}
|\tilde T_{1}^ { 1/2}> & =  &  \frac{1}{\sqrt{3}} \tilde t_{pp}^
{cR} P_{10}| \phi>
 +   \frac{1}{\sqrt{3}} \tilde t_{pp}^ {cR}   P_{10} G_0 | A'> <A' | T_0^
{\frac{1}{2}} >\cr & - &  \frac{1}{3}   \tilde t_{pp}^ {cR} P_{11}
G_0 |A' > < A'( | T_1^ {\frac{1}{2}}> + \sqrt{2}
|T_1^{\frac{3}{2}}>)\label{43}
\end{eqnarray}
with
\begin{eqnarray}
\tilde t_{pp}^{cR} \equiv t_{pp}^{cR} - | A> < A| t_{pp}^{cR}| A'>
< A'| ~. 
\label{44}
\end{eqnarray}

In (\ref{43}) the projection on partial waves to the right of
$t_{pp}^{cR}$
 has been carried out
 to infinite order  leading to the 3-dimensional
 screened Coulomb t-matrix $ t_{pp}^ {cR} $. But due to (\ref{44})
also the partial wave projected $ t_{pp}^ {cR} $ matrix occurs.
Now we insert (\ref{43})
 into the set (\ref{37}), (\ref{40}) and (\ref{41}):
\begin{eqnarray}
< A |T_0^ { 1/2}> &=& - \frac {1} {2} < A | t_{np}^{00}  P_{00}| \phi>\cr &-
& \frac {1} {2}  <A  | t_{np}^{00}  P_{00}G_0 |A'>  < A' | T_{0}^ { 1/2}>
\cr & - & \frac{\sqrt{3}}{2}  < A  | t_{np}^{00} P_{01} G_0   |A'
> < A' | T_{1}^ { 1/2}>\cr & - &\frac{\sqrt{3}}{2}  <  A |
t_{np}^{00} P_{01} G_0 | \tilde T_1^ {1/2} >  ~,
\label{45}
\end{eqnarray}
\begin{eqnarray}
< A | T_{1}^ { 1/2}>  &=&  \frac{1}{2 \sqrt{3}} < A |( t_{np}^{10}
+ 2 t_{pp}^{1-1})  P_{10} | \phi>\cr & + & \frac{1}{2 \sqrt{3}}  <
A  |(  t_{np}^{10} + 2 t_{pp}^{1 -1})   P_{10} G_0  |A'>  < A' |
T_{0}^ { 1/2}>\cr & - & \frac {1} {6}  < A |( t_{np}^{10} + 2 t_{pp}^{1 -1})
P_{11} G_0  |A'> < A' | T_{1}^ { 1/2}> \cr & + &
\frac{\sqrt{2}}{3}  < A  | (t_{np}^{10} -t_{pp}^{1 -1})  P_{11}G_0
| A'> < A' | T_{1}^ { 3/2}>\cr & - & \frac{1}{2} < A | t_{np}^{10}
P_{11}G_0  | \tilde T_{1}^ { 1/2}> ~,
\label{46}
\end{eqnarray}
\begin{eqnarray}
<A | T_{1}^ { 3/2}>  &=& \frac{1}{\sqrt{6}} <A | (t_{np}^{10}
-t_{pp}^{1 -1}) P_{10} | \phi>\cr & + &\frac{1}{\sqrt{6}}  <A |
(t_{np}^{10} -t_{pp}^{1 -1}) P_{10}G_0  | A'> < A'|T_{0}^ {
1/2}>\cr & - & \frac{ 1}{3\sqrt{2}}  <A | (t_{np}^{10} -t_{pp}^{1
-1}) P_{11}G_0  | A' > < A' |T_{1}^ {1/2}>\cr & + & \frac {1} {3}  < A | (2
t_{np}^{10} +t_{pp}^{1 -1}) P_{11}G_0 |A'> <A' |T_{1}^ { 3/2}>\cr
& - &  \frac{ 1}{\sqrt{2}}   <A |  t_{np}^{10}   P_{11}G_0 |\tilde
T_{1}^ { 1/2}> ~.
\label{47}
\end{eqnarray}

Eqs. (\ref{45}) - (\ref{47}) together with (\ref{43}) form a
closed set for amplitudes projected
 on low partial waves only. Since we anyhow iterate the equations and sum up
the resulting series
 by Pade one can stay with that form. However, we prefer to eliminate
(\ref{43})  to arrive
 at our final form of coupled equations:
\begin{eqnarray}
<A |T_0^ { 1/2}> &=& - \frac {1}{2} <A  | t_{np}^{00}  P_{00}| \phi>\cr
 & - &\frac{1}{2}  < A | t_{np}^{00} P_{01} G_0 \tilde t_{pp}^ {cR}
P_{10}|\phi>\cr
 &- & \frac {1}{2} < A | t_{np}^{00} P_{00}G_0 |A'> <A' | T_{0}^ {1/2}> \cr
 & - &\frac{1}{2}  < A  | t_{np}^{00} P_{01} G_0
\tilde t_{pp}^ {cR}  P_{10}  G_0 |A'> < A' | T_0^ {\frac{1}{2}}>\cr
 & - & \frac{\sqrt{3}}{2}  < A | t_{np}^{00} P_{01} G_0   | A'> <
A' | T_{1}^ { 1/2}>\cr
 & + &\frac{1}{2\sqrt{3}}  < A  | t_{np}^{00} P_{01} G_0  \tilde t_{pp}^
{cR} P_{11} G_0 |A' >
  < A'  | T_1^ {\frac{1}{2}}>\cr
 & + &\frac{1}{\sqrt{6}} < A | t_{np}^{00} P_{01} G_0 \tilde t_{pp}^ {cR}
P_{11} G_0 |A' >
 < A' |T_1^{\frac{3}{2}}> ~,
\label{48}
\end{eqnarray}
\begin{eqnarray}
< A | T_{1}^ { 1/2}>  &=&  \frac{1}{2 \sqrt{3}} < A |( t_{np}^{10}
+ 2 t_{pp}^{1-1})  P_{10} | \phi>\cr
 & - & \frac{1}{2\sqrt{3}} < A | t_{np}^{10}  P_{11}G_0  \tilde t_{pp}^ {cR}
P_{10} | \phi>\cr
 & + & \frac{1}{2 \sqrt{3}} < A |( t_{np}^{10} + 2 t_{pp}^{1 -1}) P_{10} G_0
|A' > < A'| T_{0}^ { 1/2}>\cr
& - & \frac{1}{2\sqrt{3}} < A  | t_{np}^{10}  P_{11}G_0 \tilde
t_{pp}^ {cR} P_{10}  G_0 | A'>  <A'| T_0^ {\frac{1}{2}} >\cr
 & - & \frac {1} {6}  < A |( t_{np}^{10} + 2 t_{pp}^{1 -1}) P_{11} 
G_0  |A'> < A' |
T_{1}^ { 1/2}> \cr
 & + & \frac{1}{6} < A | t_{np}^{10} P_{11}G_0 \tilde t_{pp}^ {cR} P_{11}
G_0 |A' > < A'  | T_1^ {\frac{1}{2}}>\cr
  & + & \frac{\sqrt{2}}{3}  <A  | (t_{np}^{10} -t_{pp}^{1 -1})  P_{11}G_0 |
A' > < A' | T_{1}^ { 3/2}>\cr
 & + & \frac{\sqrt{2}}{6} < A  | t_{np}^{10}  P_{11}G_0 \tilde t_{pp}^ {cR}
P_{11} G_0 |A' > < A' | T_1^{\frac{3}{2}}> ~,
\label{49}
\end{eqnarray}
\begin{eqnarray}
<A | T_{1}^ { 3/2}>  &=& \frac{1}{\sqrt{6}}< A  | (t_{np}^{10}
-t_{pp}^{1 -1}) P_{10} | \phi>\cr
 & - &  \frac{ 1}{\sqrt{6}} < A | t_{np}^{10}   P_{11}G_0  \tilde t_{pp}^
{cR} P_{10}| \phi>\cr
 & + & \frac{1}{\sqrt{6}} < A | (t_{np}^{10} -t_{pp}^{1 -1})P_{10}G_0 | A'>
< A' |T_{0}^ { 1/2}>\cr
 & - & \frac{
1}{\sqrt{6}} < A | t_{np}^{10}   P_{11}G_0 \tilde t_{pp}^ {cR}
P_{10} G_0 |A'> <A' | T_0^ {\frac{1}{2}} >\cr
 & - &
\frac{1}{3\sqrt{2}} <A |(t_{np}^{10} -t_{pp}^{1 -1}) P_{11}G_0 |A'
> < A' |T_{1}^ {1/2}>\cr
 & + & \frac{ 1}{3 \sqrt{2}} < A |
t_{np}^{10} P_{11}G_0 \tilde t_{pp}^ {cR} P_{11} G_0 | A' > <A'  |
T_1^ {\frac{1}{2}}>\cr
 & + &  \frac{ 1}{3} < A | (2 t_{np}^{10} +t_{pp}^{1
-1})P_{11}G_0 | A'>< A' |T_{1}^ { 3/2}>\cr
 & + & \frac{ 1}{3} < A
| t_{np}^{10}  P_{11}G_0 \tilde t_{pp}^ {cR} P_{11} G_0 | A' > <A'
|T_1^{\frac{3}{2}}> ~.
 \label{50}
\end{eqnarray}

Note that in all three equations there occur terms  where the full
3-dimensional screened Coulomb
 t-matrix $ t_{pp}^ {cR}$ is sandwiched between permutation operators and
occurs together with the strong np t-matrix. Due to (47) there are
also other terms of that second order type in t-operators where,
 however, $ t_{pp}^ {cR} $ is projected onto the lower partial waves states.

Whereas all the other expressions are of our  standard type~\cite{physrep96}
the ones with the 3-dimensional
 $t_{pp}^ {cR}$ operator require special care and will be dealt with in
appendices.

Now we would like to write down equations (\ref{48}), (\ref{49})
and (\ref{50})
 directly in our standard momentum space partial wave basis $|pq\alpha>$
which is an extension of $|pqa>$ by adding isospin quantum numbers
\begin{eqnarray}
|p q \alpha> \equiv |pqa> |(t \frac {1} {2})T> = |pq(ls)j(\lambda
\frac {1} {2})I (jI)J (t \frac {1} {2})T>\label{51} ~.
\end{eqnarray}
Though this appears as a repetition it is a necessity to cast the
set into our standard form, which underlies our existing codes.

We distinguish between the partial wave states $|pq\alpha>$ with
total 2N angular momentum $j$ below some value $j_{max}$: $j \le
j_{max}$,
 in which the nuclear, $V_N$, as well as the pp screened Coulomb
interaction, $V_c^R$
 (in isospin $t=1$ states only), are acting,  and the
states $|pq\beta>$ with $j > j_{max}$, for which only $V_c^R$ is
acting in the pp subsystem. The states $|pq\alpha>$ and
$|pq\beta>$ form a complete system of states
\begin{equation}
\int {p^2 dpq^2 dq} (\sum\limits_\alpha  {\left| {pq\alpha }
\right\rangle } \left\langle {pq\alpha } \right| +
\sum\limits_\beta  {\left| {pq\beta } \right\rangle } \left\langle
{pq\beta } \right|) = {\rm I} \\ .~ \label{52}
\end{equation}
Projecting Eq.(\ref{13}) for $ T| \Phi >$ on the $|pq\alpha>$ and
$|pq\beta>$ states one gets the following system of coupled
integral equations
\begin{eqnarray}
 \left\langle {pq\alpha } \right|T\left| {\Phi } \right\rangle  &=&
 \left\langle {pq\alpha } \right|t_{N + c}^R P\left| {\Phi } \right\rangle
\cr
 &+& \left\langle {pq\alpha } \right|t_{N + c}^R PG_0 \sum\limits_{\alpha '}
 {\int {p'^2 dp'q'^2 dq'\left| {p'q'\alpha '} \right\rangle \left\langle
 {p'q'\alpha '} \right|} } T\left| {\Phi } \right\rangle \cr
 &+& \left\langle {pq\alpha } \right|t_{N + c}^R PG_0 \sum\limits_{\beta '}
 {\int {p'^2 dp'q'^2 dq'\left| {p'q'\beta '} \right\rangle \left\langle
 {p'q'\beta '} \right|} } T\left| {\Phi } \right\rangle  \label{53} \\
 \left\langle {pq\beta } \right|T\left| {\Phi } \right\rangle  &=&
 \left\langle {pq\beta } \right|t_c^R P\left| {\Phi } \right\rangle \cr
  &+& \left\langle {pq\beta } \right|t_c^R PG_0 \sum\limits_{\alpha '}
  {\int {p'^2 dp'q'^2 dq'\left| {p'q'\alpha '} \right\rangle
  \left\langle {p'q'\alpha '} \right|} } T\left| {\Phi } \right\rangle \cr
  &+& \left\langle {pq\beta } \right|t_c^R PG_0 \sum\limits_{\beta '}
  {\int {p'^2 dp'q'^2 dq'\left| {p'q'\beta '} \right\rangle
  \left\langle {p'q'\beta '} \right|} } T\left| {\Phi } \right\rangle
 \label{54}
\end{eqnarray}
where $t_{N+c}^R$ and $t_c^R$ are t-matrices generated by
the interactions $V_N+V_c^R$ and $V_c^R$, respectively.

The third term  on the right hand side of (\ref{54}) is
proportional to $<pq\beta|t_c^RPG_0|p'q'\beta'><p'q'\beta'|t_c^ R$. A direct
calculation of its isospin part shows that independently from the
value of the total isospin $T$ it vanishes. This corresponds to
the result found in (\ref{32}).

Inserting $<pq\beta|T|\Phi>$ from (\ref{54}) into (\ref{53}) one
gets
\begin{eqnarray}
 \left\langle {pq\alpha } \right|T\left| {\Phi } \right\rangle  &=&
 \left\langle {pq\alpha } \right|t_{N + c}^R P\left| {\Phi } \right\rangle
 + \left\langle {pq\alpha } \right|t_{N + c}^R PG_0 t_c^R P\left| {\Phi }
\right\rangle \cr
  &-& \left\langle {pq\alpha } \right|t_{N + c}^R PG_0 \sum\limits_{\alpha
'}
  {\int {p'^2 dp'q'^2 dq'\left| {p'q'\alpha '} \right\rangle
  \left\langle {p'q'\alpha '} \right|} } t_c^R P\left| {\Phi } \right\rangle
\cr
  &+& \left\langle {pq\alpha } \right|t_{N + c}^R PG_0 \sum\limits_{\alpha
'}
  {\int {p'^2 dp'q'^2 dq'\left| {p'q'\alpha '} \right\rangle
  \left\langle {p'q'\alpha '} \right|} } T\left| {\Phi }
  \right\rangle \cr
  &+& \left\langle {pq\alpha } \right|t_{N + c}^R PG_0 t_c^R PG_0
\sum\limits_{\alpha '}
  {\int {p'^2 dp'q'^2 dq'\left| {p'q'\alpha '} \right\rangle
  \left\langle {p'q'\alpha '} \right|} } T\left| {\Phi } \right\rangle \cr
  &-& \left\langle {pq\alpha } \right|t_{N + c}^R PG_0  \sum\limits_{\alpha
'}
  {\int {p'^2 dp'q'^2 dq'\left| {p'q'\alpha '} \right\rangle
  \left\langle {p'q'\alpha '} \right|} } t_c^R PG_0 \cr
  && \sum\limits_{\alpha'' }
  {\int {p''^2 dp''q''^2 dq''\left| {p''q''\alpha'' } \right\rangle
  \left\langle {p''q''\alpha'' } \right|} } T\left| {\Phi } \right\rangle ~.
  \label{55}
 \end{eqnarray}
We used again a relation corresponding to (\ref{44}), now in
isospin notation.
This is a coupled set of integral equations in the space of only the states
$|\alpha>$, which incorporates the contributions of
the pp Coulomb interaction from all partial wave states up to
infinity. Using the definition of the  $|\gamma>_i$ states from
(\ref{10}) a direct calculation shows that the set (\ref{55}) is
identical to  the set (\ref{48}), (\ref{49}) and (\ref{50}).
 It can be solved by iteration and Pade
summation. There are two new leading terms
$<pq\alpha|t_{N+c}^RPG_0t_c^RP|\Phi>$  and
-$<pq\alpha|t_{N+c}^RPG_0|\alpha'><\alpha'|t_c^RP|\Phi>$. The
first term must be calculated using directly
the $3$-dimensional screened Coulomb t-matrix $t_c^R$, while the
second term requires partial wave projected screened Coulomb
t-matrix elements in the  $|\alpha>$ channels only. The kernel also contains
two new terms.  The term
$<pq\alpha|t_{N+c}^RPG_0t_c^RPG_0|\alpha'><\alpha'|T|\Phi>$ must again be
calculated with a 3-dimensional screened Coulomb t-matrix while the second
one,
-$<pq\alpha|t_{N+c}^RPG_0|
\alpha'><\alpha'|t_c^RPG_0|\alpha''><\alpha''|T|\Phi>$,  involves only
 the partial wave projected screened Coulomb t-matrix elements in the
$|\alpha>$ channels. The calculation of the new terms with the
partial wave projected Coulomb t-matrices follows our standard
procedure. Namely the  two sub kernels $t_{N+c}^RPG_0$ and
$t_c^RPG_0$ are applied consecutively  on the corresponding state.
The details how to calculate the new terms with the  3-dimensional
screened Coulomb t-matrix  are given in Appendix~\ref{a1}.

Now we turn to the physical scattering amplitudes.

\section{ The elastic pd on-shell transition amplitude}
\label{ampli}

The transition amplitude for elastic scattering is  given
by~\cite{gloeckle83,physrep96}
\begin{eqnarray}
 \left\langle {\Phi '} \right|U\left| {\Phi } \right\rangle
 &=&
 \left\langle {\Phi '} \right|PG_0^{ - 1}  + PT\left| {\Phi }
 \right\rangle ~.
\label{56}
 \end{eqnarray}
The amplitude for elastic scattering given in (\ref{56}) has two
contributions. The first one is independent of  the pp Coulomb
force
\begin{eqnarray}
< \Phi'| P G_0^{-1} | \Phi> = < \Phi'| P V | \Phi> = - \frac{1}{2}
< \phi'| P_{00}  V_{np}^{00} | \phi> ~.
\label{67}
\end{eqnarray}
For the second one we use (\ref{17}), (23) and (24)  and
decompose into lower and higher partial waves which yields using
(46)
\begin{eqnarray}
< \Phi'| P T | \Phi> & = &   - \frac{1}{2}  < \phi'| P_{00}|
T_0^{1/2} > - \frac{\sqrt{3}}{2} < \phi'| P_{01}| T_1^{1/2} >\cr
 & = & - \frac{1}{2}  < \phi'| P_{00}| A> < A | T_0^{1/2}> -
\frac{\sqrt{3}}{2} < \phi'| P_{01}|A> <A |  T_1^{1/2} >  -
\frac{\sqrt{3}}{2} < \phi'| P_{01}|\tilde  T_1^{1/2} >\cr
 & = &  - \frac{1}{2} < \phi'| P_{01}\tilde t_{pp}^{cR}P_{10} | \phi> -
\frac{1}{2}< \phi'| P_{00}|A> <A |  T_0^{1/2} >\cr & -&
\frac{1}{2}< \phi'| P_{01}\tilde t_{pp}^{cR}P_{10}G_0 | A > < A
|T_0^{1/2} > -  \frac{\sqrt{3}}{2} < \phi'| P_{01}| A > < A
|T_1^{1/2} >\cr
 & + &  \frac{1}{2 \sqrt{3}}< \phi'| P_{01}\tilde
t_{pp}^{cR}P_{11}G_0 |A> < A | (| T_1^{1/2}> + \sqrt{2} |
T_1^{3/2}>) ~.
\label{68}
\end{eqnarray}
Again besides standard low partial wave contributions terms occur
with the 3-dimensional screened Coulomb t-matrix sandwiched
between permutation operators.

To calculate the matrix element (\ref{56}) one needs
$\left\langle {\vec p\vec q~} \right|T\left| {\Phi }
\right\rangle$ composed of low and high partial wave 
contributions for $ T | \Phi>$ (see (57)). Using the completeness relation
(\ref{52}) one gets:
\begin{eqnarray}
&&\left\langle {\vec p\vec q~} \right|T\left| {\Phi }
\right\rangle  = \left\langle {\vec p\vec q~}
\right|\sum\limits_{\alpha '} {\int {p'^2 dp'q'^2 dq'\left|
{p'q'\alpha '} \right\rangle \left\langle {p'q'\alpha '} \right|}
} T\left| {\Phi } \right\rangle \cr && - \left\langle {\vec
p\vec q~} \right|\sum\limits_{\alpha '} {\int {p'^2 dp'q'^2
dq'\left| {p'q'\alpha '} \right\rangle \left\langle {p'q'\alpha '}
\right|} } t_c^R P\left| {\Phi } \right\rangle \cr &&-
\left\langle {\vec p\vec q~} \right|\sum\limits_{\alpha '} {\int
{p'^2 dp'q'^2 dq'\left| {p'q'\alpha '} \right\rangle \left\langle
{p'q'\alpha '} \right|} } t_c^R PG_0
 \sum\limits_{\alpha
''} {\int {p''^2 dp''q''^2 dq''\left| {p''q''\alpha ''}
\right\rangle \left\langle {p''q''\alpha ''} \right|} } T\left|
{\Phi } \right\rangle \cr && + \left\langle {\vec p\vec q~}
\right|t_c^R P\left| {\Phi } \right\rangle  + \left\langle
{\vec p\vec q~} \right|t_c^R PG_0 \sum\limits_{\alpha '} {\int
{p'^2 dp'q'^2 dq'\left| {p'q'\alpha '} \right\rangle \left\langle
{p'q'\alpha '} \right|} } T\left| {\Phi } \right\rangle ~.
\label{69}
\end{eqnarray}

It follows, that in addition to the amplitudes $<pq\alpha|T|\Phi >$ also the
partial wave projected amplitudes $<pq\alpha|t_c^RP|\Phi >$ and
$<pq\alpha|t_c^RPG_0|\alpha'><\alpha'|T|\Phi>$ are required. The
expressions for the contributions of these three terms to the
transition amplitude for the elastic scattering (and the breakup reaction) are
given in Appendix \ref{b1}.
The last two terms in (\ref{69}) must  again  be calculated using
directly the $3$-dimensional screened Coulomb t-matrices. In Appendix
\ref{c1} the expressions for $\left\langle {\vec p\vec q~}
\right|t_c^R P\left| {\Phi } \right\rangle$ (breakup) and
$\left\langle {\vec p\vec q~} \right|Pt_c^R P\left| {\Phi }
\right\rangle$ (elastic scattering) are given together with
the expression for $<\Phi'|PG_0^{-1}|\Phi>$. In Appendix  \ref{d1} we
describe how to get the last matrix element $<\vec p \vec q~
|t_c^RPG_0|\alpha '><\alpha '|T|\Phi>$.

\section{The screening limit}
\label{limit}

The set of coupled Faddeev equations (\ref{55}) or (\ref{45})-(\ref{47})
 is well defined for a finite screening radius. It is
an exact set assuming that the strong NN t-matrix can be neglected
beyond a certain $ j_{max}$, which is   justified. Further the pp
screened Coulomb force is taken  into account to infinite order in
the partial wave decomposition in form of the 3-dimensional
screened Coulomb t-matrix $ t_{pp}^{cR}$. The important challenge
is to control the screening  limit for the physical on shell
amplitude (\ref{56}). The
contribution (\ref{67}) is well defined and independent of the
Coulomb force. The corresponding expression  without partial-wave expansion
is given in
(\ref{eq.a2.5}).

In the second part given in (\ref{68}) there are
contributions without and with an explicit $ \tilde t_{pp}^{cR}$
operator. The first term in (\ref{68}) is responsible for the
Rutherford scattering between proton and deuteron. It has the form
\begin{eqnarray}
< \phi'| P_{01}\tilde t_{pp}^{cR} P_{10}| \phi> & = &  < \phi'|
P_{01} t_{pp}^{cR} P_{10}| \phi>  - < \phi'| P_{01}| A> < A
|t_{pp}^{cR}| A'> < A'| P_{10}| \phi> ~.
\label{70}
\end{eqnarray}
With $ P_{01} = P_{10} = P_{12} P_{23} - P_{13} P_{23} $ and using
simple symmetry properties one obtains
\begin{eqnarray}
< \phi'| P_{01} t_{pp}^{cR} P_{10}| \phi> & = &  2< \phi'| P_{12}
P_{23}  t_{pp}^{cR}( P_{12} P_{23} - P_{13} P_{23}) |\phi> ~.
\label{71}
\end{eqnarray}
The 3-dimensional screened Coulomb t-matrix enters that matrix
element totally off-shell, except for forward scattering. To show
that it is sufficient to regard only the momentum space part of
that matrix element. Using well known relations among Jacobi
momenta and the symmetry property of the deuteron one obtains
\begin{eqnarray}
& & < \phi'| P_{01} t_{pp}^{cR} P_{10}| \phi>  \sim   2 \int d^3
q' \phi_d( 1/2 \vec q_0~' + \vec q~')  \phi_d( 1/2 \vec q_0 + \vec
q~')\cr
 & &  (t_{pp}^{cR} ( - \vec q_0~' - 1/2 \vec q~', 1/2 \vec q~'
+ \vec q_0, E( q')) -t_{pp}^{cR} ( - \vec q_0~' - 1/2 \vec q~',- 1/2
\vec q~' - \vec q_0, E( q'))
\label{72}
\end{eqnarray}
with
\begin{eqnarray}
E( q') = E- \frac{3}{4m} q'^2 = E_d + \frac{3}{4m} q_0^2
-\frac{3}{4m} q'^2 ~.
\label{73}
\end{eqnarray}
It is easy to see that $(\vec q_0~' + 1/2 \vec q~')^2 /m \ne E(q')
\ne(\vec q_0 + 1/2 \vec q~')^2 /m $ and obviously $ (\vec q_0~' +
1/2 \vec q~')^2 /m \ne (\vec q_0 + 1/2 \vec q~')^2 /m $ except for
forward scattering. Now the off-shell screened Coulomb t-matrix
converges for $ R \rightarrow \infty $ towards the pure off-shell
Coulomb t-matrix, which is well defined. Thus except for forward
scattering that matrix element has a well defined screening limit.

It is interesting to regard the replacement $ t_{pp}^{cR} \rightarrow
V_{pp}^{cR} \equiv \frac{e^2}{r} e^{-r/R} $ which in the limit $ R
\rightarrow \infty $ turns (\ref{72})
into
\begin{eqnarray}
< \phi'| P_{01} V_{pp}^{cR} P_{10}| \phi> &\sim&
 \frac{- e^2}{ ( \vec q_0 - \vec q_0~')^2 } \int d^3 q'\phi_d( 1/2
\vec q_0~' + \vec q~')  \phi_d( 1/2 \vec q_0 + \vec q~')\cr & -&\int
d^3 q'\phi_d( 1/2 \vec q_0~' + \vec q~')  \phi_d( 1/2 \vec q_0 +
\vec q~')\frac{e^2}{ ( \vec q_0 + \vec q_0~' + \vec q~')^2} ~.
\label{74}
\end{eqnarray}
The first part is the Rutherford scattering  amplitude multiplied
by a deuteron form factor and the second term a correction due to
antisymmetrisation of the two protons.

The partial wave projected piece in (\ref{70}) has also a well
defined screening limit due to the same reasons.

Next we regard the two matrix elements of $\tilde t_{pp}^{cR} $
 in (61) integrated together with $ |A>$-projected $ T$-amplitudes.
  As  before $t_{pp}^{cR}$ is
 off-shell from the left. On the right there is an intermediate momentum
integration, which includes the
on-shell point.
  In that case $t_{pp}^{cR}$  acquires an infinitely oscillating phase
 factor in the screening limit~\cite{chen72}
  which, however, is integrable.
 Thus it remains to verify that the free propagator singularity of $ G_0$
does
 not coincide with that on-shell   singularity of $ t_{pp}^{cR}$. Inserting
 the momentum representation and working out the
 permutations the free propagator singularity is located at
\begin{eqnarray}
E_d + \frac{3}{4m} q_0^2 - \frac{( \vec q~' + 1/2 \vec q~)^2 }{m}
- \frac{3}{4m} q^2 =0 \label{75}
\end{eqnarray}
where $ \vec q~' $ and $ \vec q$ are integration variables. On the
other hand the singularity of $ t_{pp}^{cR}( \vec p,\vec p~',E(q'))
$ arising for $ \vec p - \vec p~' =0$ is located in the matrix
element either  at $ \vec q_0~' + \vec q~' + \vec q =0$ or at $\vec
q_0~' = \vec q$. This inserted into (\ref{75}) yields in both cases
a non vanishing expression. Thus 
 in that respect all the three terms in (\ref{68})
containing $t_{pp}^{cR} $ 
 have a well defined screening limit. 
 It remains to consider
 the partial wave projected T-amplitudes $ < A| T_t^T>$,  
 which enter in the matrix elements in (\ref{68}). 
 They are the solutions of
the coupled set of Faddeev equations (\ref{55}) which are ill
defined in the screening limit. However, we  expect that 
in that limit these amplitudes $< A| T_t^T> $ acquire only
infinitely oscillating logarithmic phase factors, which can 
well be integrated over. In fact as will be demonstrated in a forthcoming 
paper only the on-shell amplitude $ < A| T_t^T>$ with 
$p^2 + \frac {3} {4} q^2 = mE$, which appears in the breakup 
transition amplitude, acquires that oscillating factor.

Thus we come to the conjecture that the physical on-shell elastic
pd amplitude has  
 a well defined screening limit and does not
require renormalisation. This might appear surprising at a first
glance, since the on-shell pp scattering amplitude has not a well
defined screening limit. Our explanation is that in the pd system
the Coulomb force does not act between a proton and the center of mass of the
deuteron but
between the two protons, where one of them is part of the deuteron and
therefore the 
 pp Coulomb t-matrix  is integrated  over 
 the deuteron wave
function in the final state in the elastic scattering amplitude. 

That analytical insight is well supported
by our numerical results
laid out in the following section.

The case of the pd breakup process is quite
different and will be dealt with in a forthcoming study.

\section{Numerical results}
\label{results}

To demonstrate the feasibility of our approach we applied the outlined
formalism to a simple dynamical model in which the nucleon-nucleon
force was restricted
to act in $^1S_0$ and $^3S_1-^3D_1$ partial waves only
 and taken from the CD~Bonn potential~\cite{cdbonn}. The
 pp Coulomb force was screened exponentially
\begin{eqnarray}
V_{pp}^R(r) &=& \frac{e^2}{r} e^{-({\frac{r} {R}})^n }
\label{scr_1}
\end{eqnarray}
with the screening radius $R$ and $n=1$.

To investigate the screening limit $R \to \infty$ we generated a
set of partial-wave decomposed t-matrices, $t_c^R$, based on the
screened pp Coulomb force alone  or for $ t_{ N+C}^ R $  combined with the
pp nuclear
interaction, taking $R=20, 40, 60, 80, 100, 120$ and
$140$~fm. With that dynamical input we solved the set of Faddeev
equations (\ref{55}) for the total angular momenta of the p-p-n
system up to $J \le \frac {15} {2}$ and both parities. In this
first study we restricted ourselves to the perturbative
approximation for the 3-dimensional pure Coulomb t-matrix: $t^R_c =
V^R_c$.  Of course in the future studies that approximation will be
avoided and the full solution of the 3-dimensional LS equation for
the screened pp Coulomb t-matrix will be used.

\begin{figure}
\includegraphics[scale=1.0]{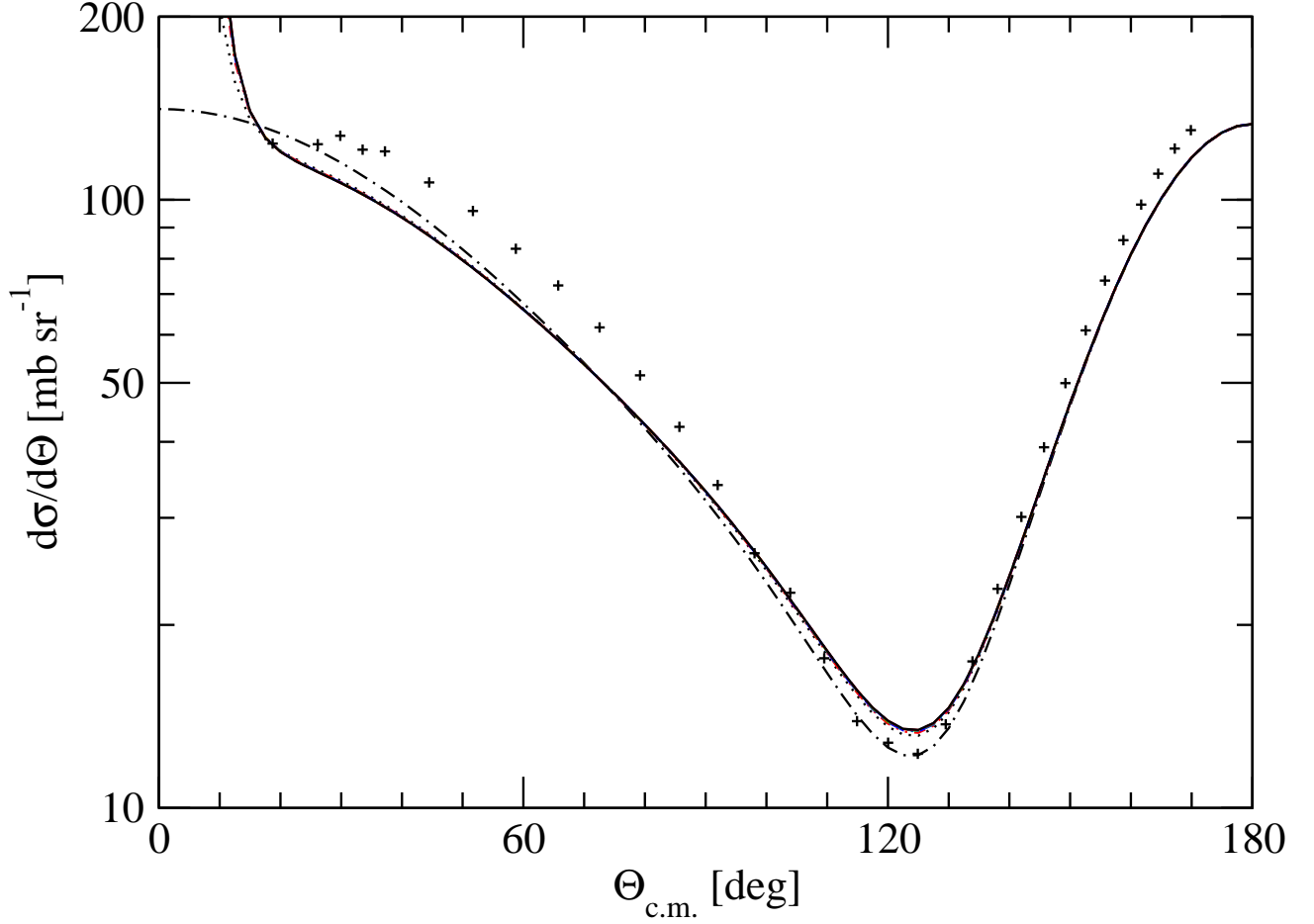}
\caption{(color online) The convergence in the cut-off radius R
of the  pd elastic scattering cross section
$\frac {d\sigma} {d\Omega}$ shown as a function of the c.m. angle
$\Theta_{c.m.}$ at the incoming proton energy
$E_p^{lab}=13$~MeV. These cross sections were
calculated with the screened Coulomb
force and the CD~Bonn nucleon-nucleon potential~\cite{cdbonn}
restricted to the $^1S_0$ and $^3S_1$-$^3D_1$ partial waves.
 The screening
 radii are :  $R=20$~fm (black dotted line),
 $R=40$~fm (green double-dashed -dotted line),
 $R=60$~fm (blue long-dashed-dotted line),
 $R=80$~fm (red dashed- double-dotted line),
 $R=100$~fm (blue short-dashed line),
 $R=120$~fm (red long-dashed line),
 $R=140$~fm (black solid line).  The $R=40$-$140$~fm lines are
practically overlapping.
The black dashed-dotted line is the corresponding
 nd elastic scattering cross section.
The pluses are $E_p^{lab}=12$~MeV pd elastic scattering cross section
data of Ref.~\cite{grub12}.}
 \label{fig1}
\end{figure}

\begin{figure}
\includegraphics[scale=1.0]{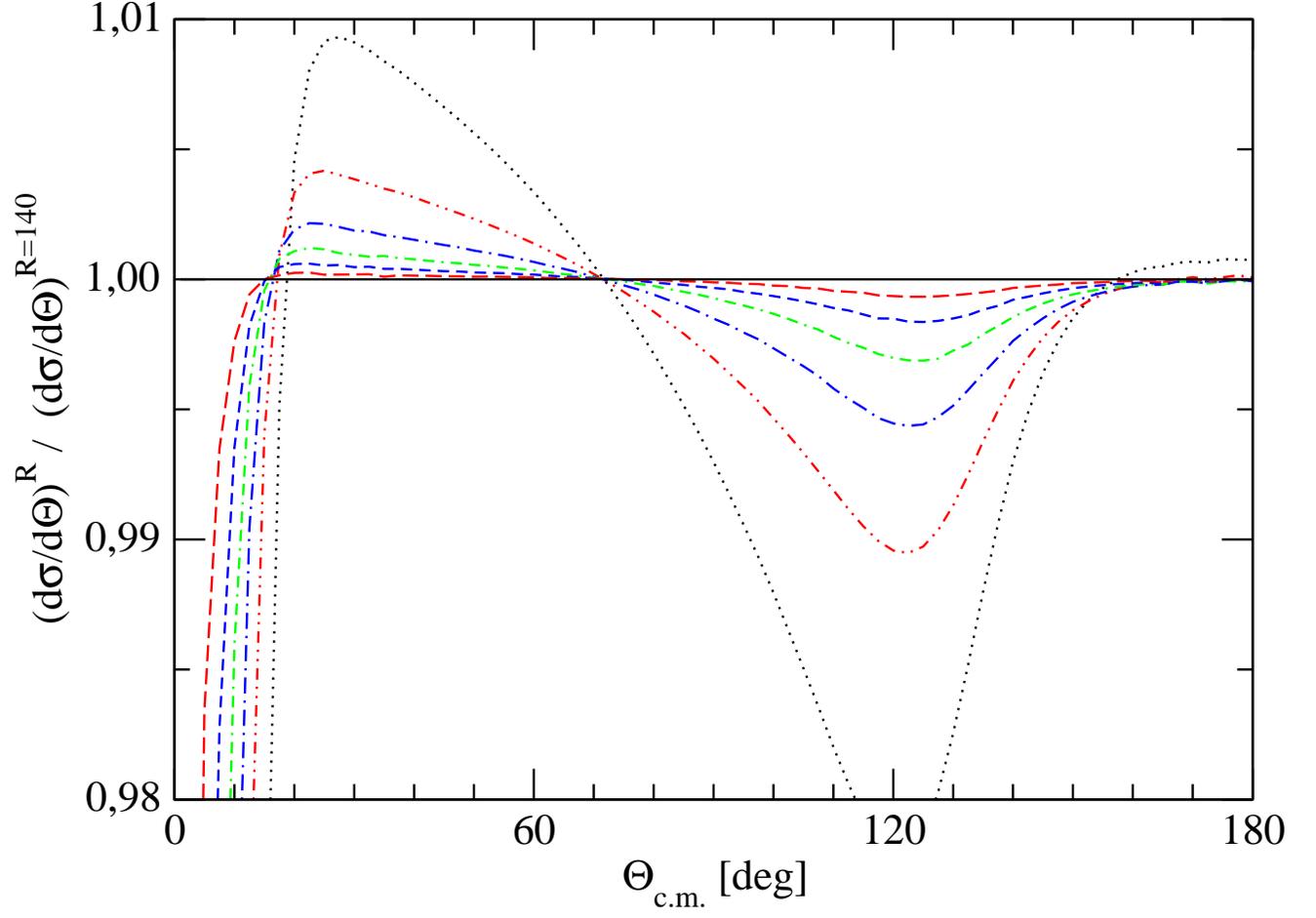}
\caption{(color online) The convergence in the cut-off radius R
of the  pd elastic
scattering cross section
 $\frac {d\sigma} {d\Omega}$ at the  incoming proton energy
$E_p^{lab}=13$~MeV, shown as the ratio
$\frac {d\sigma} {d\Omega}^R/\frac {d\sigma} {d\Omega}^{R=140}$.
For the description of the lines see Fig.~\ref{fig1}.}
 \label{fig2}
\end{figure}

In Fig.~\ref{fig1} we show the convergence in the screening radius $R$ of
the pd elastic scattering cross section and compare the pd and  nd
elastic scattering angular distributions at the incoming nucleon energy
$E_N^{lab}=13$~MeV. On the scale of the figure the cross sections for $ R=40
-140$~fm  are practically
indistinguishable. The detailed picture of that convergence is
depicted in Fig.~\ref{fig2}, where the ratio of the cross sections
obtained
with the screening radius $R$ to those with $R=140$~fm is shown as a
function of the c.m. scattering angle $\Theta_{c.m.}$. It is clearly
seen that already with the screening radius $R=40$~fm
 converged results for the cross section are achieved. Increasing
 further the value of $R$
provides cross sections which differ less than $\approx 1
\%$ up to the forward scattering angles $\Theta_{c.m.} \approx 10^o$. At
very
forward angles, where the pp Coulomb force is dominant,
 larger screening radii are required to get the cross section with the
 same precision.

\begin{figure}
\includegraphics[scale=1.0]{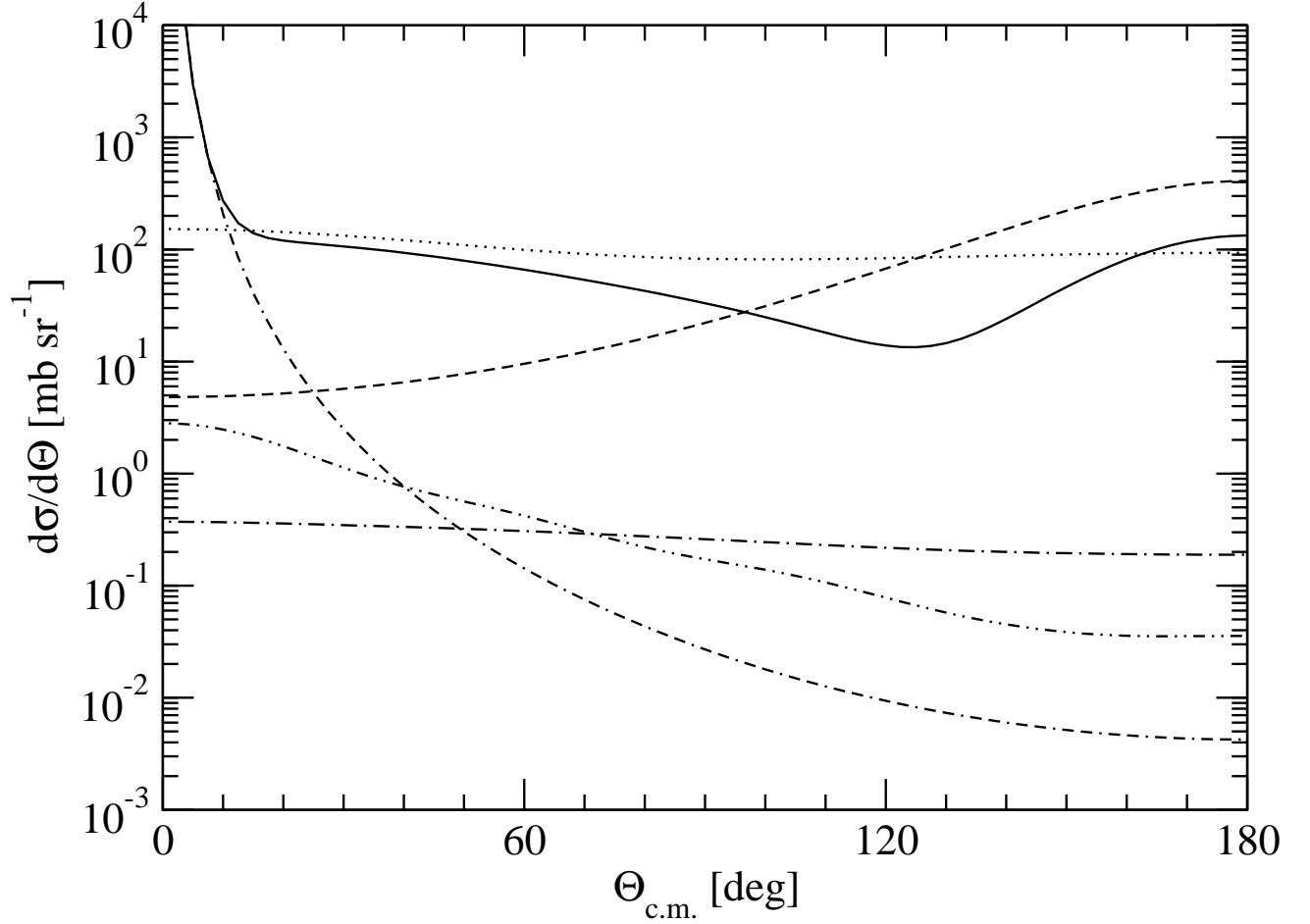}
\caption{The contributions of different terms to
 the  pd elastic
scattering cross section
 $\frac {d\sigma} {d\Omega}$ at the incoming proton energy
$E_p^{lab}=13$~MeV
calculated with the screening radius
 $R=100$~fm.
 The dotted and dashed lines are contributions of the
 $ \left\langle \Phi' |  PT | {\Phi } \right\rangle $ and
 $ \left\langle \Phi' |  PG_0^{-1} | {\Phi } \right\rangle $
terms, respectively. The double-dashed-dotted line is  the
contribution of the
$ \left\langle \Phi' |  Pt_c^RP | {\Phi } \right\rangle $
 term coming with the 3-dimensional screened
Coulomb t-matrix $t_c^R$. The dashed-double-dotted and dashed-dotted
lines are contributions of the
 $ \left\langle \Phi' |  Pt_c^RP | {\Phi } \right\rangle $ and
 $ \left\langle \Phi' |  Pt_c^RPG_0T | {\Phi } \right\rangle $
terms, respectively, which are calculated with the  partial-wave decomposed
 screened Coulomb t-matrix. The solid line encompasses all terms. In this
feasibility study the 3-dimensional $ t_c^ R$ is replaced by $ V_c^ R$.}
 \label{fig3}
\end{figure}

The angular distributions shown in Figs.~\ref{fig1} and \ref{fig2} were
obtained taking in  the elastic scattering transition amplitude
(\ref{56})  the exchange term
 $ \left\langle \Phi' |  PG_0^{-1} | {\Phi } \right\rangle $
  together with the first four terms in (\ref{69}) contributing to
 $ \left\langle \Phi' |  PT | {\Phi } \right\rangle $.
 In Fig.~\ref{fig3} we
present how each term contributes to the cross section.
 When all
 terms are taken into account the resulting angular distribution
 is given  by the solid line.
The $ \left\langle \Phi' |  PT | {\Phi } \right\rangle $ term
( dotted line related to the first term in (\ref{69})) contributes
significantly at all angles. At backward angles the
largest contribution comes from the exchange term
 $ \left\langle \Phi' |  PG_0^{-1} | {\Phi } \right\rangle $ (dashed
line)  while at forward angles the most important is the ``Rutherford''
term $ \left\langle \Phi' |  Pt_c^RP | {\Phi } \right\rangle $
(double-dashed-dotted line related to the  fourth term in (\ref{69}))
calculated with the 3-dimensional screened
Coulomb t-matrix $t_c^R$ (in this first study treated perturbatively
as $t_c^R=V_c^R$). The  two terms  based on the partial-wave
projected Coulomb t-matrix,
$ \left\langle \Phi' |  Pt_c^RP | {\Phi } \right\rangle $
(dashed-double-dotted line related to  the second term in (\ref{69}) ) and
 $ \left\langle \Phi' |  Pt_c^RPG_0T | {\Phi } \right\rangle $
(dashed-dotted line related to the third term in (\ref{69}) ), are about
$2$-orders of magnitude smaller and thus of minor importance. The fact
that at very forward angles the contribution of the
 $ \left\langle \Phi' |  Pt_c^RPG_0T | {\Phi } \right\rangle $ is an
order of magnitude smaller than the contribution of
the $ \left\langle \Phi' |  Pt_c^RP | {\Phi } \right\rangle $ seems to
justify the neglection of the last term
 $ \left\langle \Phi' |  Pt_c^RPG_0T | {\Phi } \right\rangle $ in
 (\ref{69}) coming with the 3-dimensional screened Coulomb t-matrix.
In future studies this term will  be calculated to verify
this statement.

\begin{figure}
\includegraphics[scale=0.8]{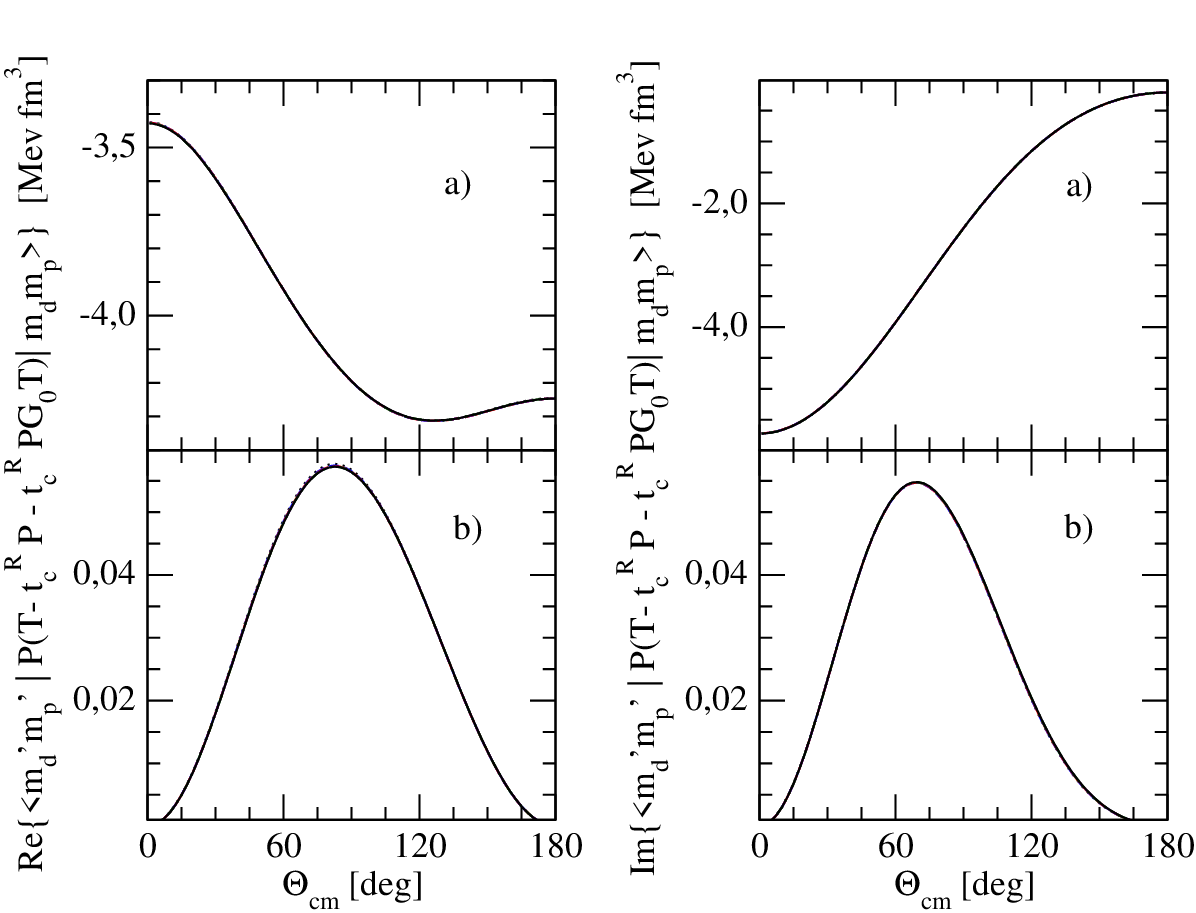}
\caption{(color online) 
The independence of the real (left column) and imaginary 
(right column) parts of the partial wave 
contribution 
 $ \left\langle \Phi'_{m_{d'}m_{p'}} |  P(T - t_c^RP - t_c^RPG_0T) 
| {\Phi_{m_{d}m_{p}} } \right\rangle $
to the $E_p^{lab}=13$~MeV pd 
elastic scattering transition amplitude 
 on the cut-off radius R. The different lines are: $R=20$~fm - dotted, 
 $R=40$~fm - short-dashed, $R=60$~fm - long-dashed, 
 $R=80$~fm - short-dashed-dotted, $R=100$~fm - long-dashed-dotted, 
 $R=120$~fm - double-dotted-dashed, $R=140$~fm - solid. All the lines are 
practically overlapping. The incoming and 
outgoing deuteron and proton spin projections are for a): $m_d=m_{d'}=-1$ and 
$m_p=m_{p'}=-\frac {1} {2}$ and for b): $m_d=-1$, $m_{d'}=+1$, 
 and $m_p=m_{p'}=+\frac {1} {2}$.}
 \label{fig4}
\end{figure}

In Fig.~\ref{fig4} we demonstrate numerically that the elastic 
pd amplitude has a well defined screening limit and does not require 
renormalization. The real and imaginary parts of the partial wave 
contribution  $ \left\langle \Phi' |  P( T 
- t_c^RP - t_c^RPG_0T) | {\Phi } \right\rangle $ to the elastic 
transition amplitude are shown for two combinations of the 
incoming and outgoing deuteron and proton spin projections and a number 
of screening radia $R=20, 40, 60, 80, 100, 120$, and $140$~fm. 
 The additional term (\ref{67}) is real and independent of the screening 
radius. The fourth term in (\ref{69}) is also real under our 
approximation $t_c^R=V_c^R$ and for angles different from zero has a well 
defined screening limit. Moreover it is peaked in forward direction and would 
dominate terms shown. 
All lines are practically overlapping. That shows that not 
only the cross section but the pd elastic scattering amplitude itself 
does not develop an oscillating phase in the infinite screening limit.

\section{Summary}
\label{sumary}

We developed and presented a novel approach to include the pp
Coulomb force into the momentum space 3N Faddeev calculations. It
is based on a standard formulation for short range forces and
relies on a screening of the long-range Coulomb interaction. In order
to avoid all uncertainties connected with an application of the
partial wave expansion, unsuitable when working with long-range
forces, we apply directly the 3-dimensional pp screened Coulomb
t-matrix. Furthermore we strictly insisted to treat the pp 
Coulomb force in its proper coordinate. 

Using a simple dynamical model for the nuclear part of the
interaction we demonstrated the feasibility of that approach. We
provided analytical arguments  and showed numerically that the
physical elastic pd scattering amplitude has a well defined
screening limit and therefore does not require renormalisation.
Well converged elastic pd cross sections have been achieved at
finite screening radii. In this first study the 3-dimensional screened pure
Coulomb t-matrix was replaced by the screened
Coulomb potential and only a small number of partial wave states
for the NN interaction was taken into account. This restriction
will be removed in a forthcoming article.

\section*{Acknowledgments}
This work was supported by the Polish 2008-2011 science funds as a
 research project No. N N202 077435.
It was also partially supported by the Helmholtz
Association through funds provided to the virtual institute ``Spin
and strong QCD''(VH-VI-231)  and by
  the European Community-Research Infrastructure
Integrating Activity
``Study of Strongly Interacting Matter'' (acronym HadronPhysics2,
Grant Agreement n. 227431)
under the Seventh Framework Programme of EU. 
 H.W. would like to thank for the warm hospitality and support during the stay
 at TUNL, USA. 
 The numerical
calculations have been performed on the
 supercomputer cluster of the JSC, J\"ulich, Germany.

\clearpage

\appendix

\section{Matrix elements with 3-dimensional screened Coulomb t-matrix:
$<pq\alpha|t_{N+c}^RPG_0t_c^RPG_0|\alpha'><\alpha'|T|\Phi >$ and
$<pq\alpha|t_{N+c}^RPG_0t_c^RP|\Phi >$}
\label{a1}

These terms must be calculated using directly the 3-dimensional
screened Coulomb t-matrix $t_c^R$.

Introducing partial wave states $|pq\beta>$ in LS-coupling
\begin{eqnarray}
\left| {pq\beta } \right\rangle  \equiv \left| {pq(l\lambda
)L(s\frac{1}{2})S(LS)J(t\frac{1}{2})T} \right\rangle \label{aa1.1}
\end{eqnarray}
and using the recoupling
\begin{eqnarray}
\left| {pq\alpha } \right\rangle  = \sum\limits_{LS} {\sqrt {\hat
j\hat I\hat L\hat S} \left\{ {\begin{array}{*{20}c}
   l & s & j  \\
   \lambda  & {\frac{1}{2}} & I  \\
   L & S & J  \\
\end{array}} \right\}\left| {pq\beta } \right\rangle  \equiv
\sum\limits_\beta  {\left\langle {\beta }
 \mathrel{\left | {\vphantom {\beta  \alpha }}
 \right. \kern-\nulldelimiterspace}
 {\alpha } \right\rangle \left| {pq\beta } \right\rangle } }
 \label{aa1.2}
\end{eqnarray}
the matrix element $ \left\langle {pq\alpha } \right|t_{N+c}^RPG_0
t_c^R PG_0 \left| {\alpha '} \right\rangle \left\langle {\alpha '}
\right|T\left| {\Phi  } \right\rangle $
 can be written as
\begin{eqnarray}
&&\left\langle {pq\alpha } \right|t_{N + c}^R PG_0 t_c^R PG_0
\left| {\alpha '} \right\rangle \left\langle {\alpha '}
\right|T\left| {\Phi } \right\rangle  = \sum\limits_{\tilde
\alpha } {\int {\tilde p^2 d\tilde p ~t_{\alpha \tilde \alpha }^{N
+ c,R} (p,\tilde p;q)G_0 (\tilde p,E(q))}} \cr
&& \sum_{\tilde\beta }
\left\langle {\tilde \alpha } | {\tilde \beta } \right\rangle
 \sum_{\alpha ',\beta '}
 \left\langle {\beta '} | {\alpha '}  \right\rangle
  \sum_{\mu ,\mu '} \left\langle L\mu SM - \mu | JM \right\rangle
 \left\langle L'\mu 'S'M - \mu ' | JM \right\rangle \cr
&& {\int {p'^2 dp'q'^2} dq'G_0
(p',q')Z(\tilde pqp'q')}
  \label{aa1.3}
\end{eqnarray}
where $E(q) \equiv E-\frac {3} {4m}q^2$, $G_0(p,q) \equiv \frac {1} {E-
\frac {p^2} {m}-\frac {3q^2}
{4m} +i\epsilon }$ and
\begin{eqnarray}
&&Z(\tilde pqp'q') \equiv \delta _{SS'} \delta _{\mu \mu '} \delta
_{M_T , - 1/2} \delta _{M_{T'} , - 1/2} 6\sqrt {\hat t\hat t'}
\left\{ {\begin{array}{*{20}c}
   {1/2}  \\
   {1/2}  \\
\end{array}\begin{array}{*{20}c}
   {1/2}  \\
   T  \\
\end{array}\begin{array}{*{20}c}
   t  \\
   1  \\
\end{array}} \right\}\left\{ {\begin{array}{*{20}c}
   {1/2}  \\
   {1/2}  \\
\end{array}\begin{array}{*{20}c}
   {1/2}  \\
   {T'}  \\
\end{array}\begin{array}{*{20}c}
   1  \\
   {t'}  \\
\end{array}} \right\} \cr
&& \left[ ( - 1)^{1 + s + t'} \sqrt {\hat s\hat s'} \left\{
{\begin{array}{*{20}c}
   {1/2}  \\
   {1/2}  \\
\end{array}\begin{array}{*{20}c}
   {1/2}  \\
   S  \\
\end{array}\begin{array}{*{20}c}
   s  \\
   {s'}   \\
\end{array}} \right\} X_1(\tilde p q \tilde {\beta}, p'q' \beta')
 + \delta _{ss'} X_2(\tilde p q \tilde {\beta}, p'q' \beta') \right] ~.
  \label{aa1.4}
\end{eqnarray}
$X_1$ and $X_2$ are given by
\begin{eqnarray}
X_1(\tilde p q \tilde {\beta}, p'q' \beta')
 = \left\langle {\tilde pq(\tilde l\lambda )L\mu }
\right|P_{12}^m P_{23}^m t_c^R (T,T')P_{12}^m P_{23}^m \left |
{p'q'(l'\lambda ')L'\mu '} \right\rangle ~,
  \label{aa1.5}
\end{eqnarray}
\begin{eqnarray}
X_2(\tilde p q \tilde {\beta}, p'q' \beta')
 = \left\langle {\tilde pq(\tilde l\lambda )L\mu }
\right|P_{12}^m P_{23}^m t_c^R (T,T')P_{13}^m P_{23}^m \left|
{p'q'(l'\lambda ')L'\mu '} \right\rangle ~,
  \label{aa1.6}
\end{eqnarray}
and the isospin matrix element of the 3-dimensional screened
Coulomb t-matrix is
\begin{eqnarray}
\left\langle {(t\frac{1}{2})Tm_T } \right|t_c^R \left|
{(t'\frac{1}{2})T'm_{T'} } \right\rangle  = \delta _{t1} \delta
_{tt'} \delta _{M_T M_{T'} } \delta _{M_{T'} , - \frac{1}{2}}
t_c^R (T,T') ~.
\label{aa1.7}
\end{eqnarray}
with
\begin{eqnarray}
 t_c^R (T = \frac{1}{2},T' = \frac{1}{2}) = \frac{2}{3}t_c^R  \\
 t_c^R (T = \frac{1}{2},T' = \frac{3}{2}) = t_c^R (T = \frac{3}{2},T' =
\frac{1}{2}) =  - \frac{{\sqrt 2 }}{3}t_c^R  \\
 t_c^R (T = \frac{3}{2},T' = \frac{3}{2}) = \frac{1}{3}t_c^R
 \end{eqnarray}
The transposition operator $P_{ij}$ has been decomposed into three
parts: $P_{ij}^t$  acting on isospin-, $P_{ij}^s$ on  spin-,  and
$P_{ij}^m$ on  momentum-components  of the basis
$|pq\alpha>$
\begin{eqnarray}
P=P_{12}^mP_{23}^mP_{12}^sP_{23}^sP_{12}^tP_{23}^t +
P_{13}^mP_{23}^mP_{13}^sP_{23}^sP_{13}^tP_{23}^t ~.
\label{aa1.8}
\end{eqnarray}
Because $X_i$ are scalars diagonal in $L$ and $\mu$ and in
addition independent on $\mu$ one can write
\begin{eqnarray}
X_1 &=& \left\langle {\tilde pq(\tilde l\lambda )L\mu }
\right|P_{12}^m P_{23}^m t_c^R (T,T')P_{12}^m P_{23}^m \left |
{p'q'(l'\lambda ')L' \mu' } \right\rangle \cr
  &=& \frac{\delta_{LL'}} {2L+1} \sum_{\mu}
  \left\langle {\tilde pq(\tilde l\lambda )L\mu }
\right|P_{12}^m P_{23}^m t_c^R (T,T')P_{12}^m P_{23}^m \left |
{p'q'(l'\lambda ')L \mu } \right\rangle ~.
  \label{aa1.9}
\end{eqnarray}
Inserting in (\ref{aa1.9}) from the left and right a complete
system of states $|\vec p \vec q>$ one gets
\begin{eqnarray}
 {\rm X}_{\rm 1}  &=& \int {dp_1 p_1^2 \int {dp_1 }~' p_1^2 ~'\int {dq_1
q_1^2
 \int {d\hat {\tilde p} \int {d\hat q\int {d\hat p~'
 \int {d\hat q~'\int {d\hat p_1 } } } } } } }
 \int {d\hat p_1 } '\int {d\hat q_1 }
 \cr
 &&\frac{{\delta _{LL'} }}{{2L + 1}}\sum\limits_\mu
  {Y_{\tilde l\lambda }^{*L\mu } (\hat {\tilde p},\hat q)
  t_c^R (T,T';p_1 ,p_1~ ',\hat p_1  \circ \hat p_1 ~';E(q_1 ))}
   Y_{l'\lambda '}^{L\mu } (\hat p~',\hat q~')
 \cr
 &&\delta (\vec {\tilde p} - \frac{1}{2}\vec q - \vec q_1 )
 \delta (\vec p_1  + \vec q + \frac{1}{2}\vec q_1 )
 \delta (\vec p_1 ~' - \frac{1}{2}\vec q_1  - \vec q~')
 \delta (\vec p~' + \vec q_1  + \frac{1}{2}\vec q ~') ~.
  \label{aa1.10}
 \end{eqnarray}
The quantity $\sum\limits_\mu
  Y_{\tilde l\lambda }^{*L\mu } (\hat {\tilde p},\hat q)
   Y_{l'\lambda '}^{L\mu } (\hat p~',\hat q~')$ is a scalar and
   therefore can only depend on scalars formed from
$\hat {\tilde p}$, $\hat q$, $\hat p ~'$, and $\hat q ~'$. Only 5
of the  possible scalars are independent e.q. $\hat {\tilde p}  \circ
\hat q$, $\hat {\tilde p}  \circ \hat p~'$,  $\hat {\tilde p}
\circ \hat pq~'$, $\hat q  \circ \hat p~'$, and $\hat q \circ \hat
q~'$, and  they determine the geometry of $\hat {\tilde p}$, $\hat
q$, $\hat p~'$ and $\hat q~'$. Using the $\delta$-functions in
(\ref{aa1.10}) it  follows that
\begin{eqnarray}
 \vec {\tilde p} &=& \frac{1}{2}\vec q + \vec q_1  \cr
 \vec p_1  &=&  - \vec q - \frac{1}{2}\vec q_1  \cr
 \vec p_1 ~' &=& \frac{1}{2}\vec q_1  + \vec q~' \cr
 \vec p~' &=&  - \vec q_1  - \frac{1}{2}\vec q~' ~.
 \label{aa1.11}
 \end{eqnarray}
The above 5 scalar products can be expressed by the only
independent scalar $\vec p_1 \circ \vec p_1~'$ and by the
magnitudes of seven vectors: $\vec {\tilde p}$, $\vec q$, $\vec
p~'$, $\vec q~'$, $\vec p_1$, $\vec q_1$ and $\vec p_1~'$. This
together with (\ref{aa1.11}) enables  to perform the integrations
in (\ref{aa1.10}) by choosing the vector $\vec q_1\left\| {\hat z}
\right.$ and $\vec q \in (x - z)$. Such choice of coordinates
makes that also $\vec {\tilde p}$ and $\vec p_1 \in (x - z)$ and:
$\phi _{{\rm \tilde p}}  = 0$, $\phi _{{\rm p}_{\rm 1} } = \pi$, $ \phi
_{{\rm p}_{\rm 1} '}  = \phi _{q'}$, 
 $\phi _{{\rm p}'}  = \phi _{q'}  + \pi$.

As a result of the integration one gets
\begin{eqnarray}
 {\rm X}_{\rm 1} (\tilde p q \tilde \beta, p' q' \beta ')
  &=& \frac{{32\pi ^2 }}{{\tilde pp'qq'}}
\int\limits_{q_1^{\min } }^{q_1^{\max } }
 {dq_1 \frac{{\delta _{LL'} }}{{2L + 1}}\sum\limits_\mu
 {Y_{\tilde l\lambda }^{*L\mu } (\theta _{\tilde p} ,
 \phi _{\tilde p}  = 0,\theta _q ,\phi _q  = 0)} }  \cr
  &\times& \int\limits_0^{2\pi } {d\phi _{q'} t_c^R (T,T';
  \sqrt {\tilde p^2  + \frac{3}{4}q^2  - \frac{3}{4}q_1^2 } ,
  \sqrt {p~'^2  + \frac{3}{4}q~'^2  - \frac{3}{4}q_1^2 } ,
  \hat p_1  \circ \hat p_1 ~',E(q_1 ))}  \cr
  &\times& Y_{l'\lambda '}^{L\mu } (\theta _{p~'} ,\phi _{p~'}
   = \phi _{q '}  + \pi ,\theta _{q~'} ,\phi _{q~'} )
   \label{aa1.12}
 \end{eqnarray}
with $\hat p_1 \circ \hat p_1' = cos\theta_{p_1}cos\theta_{p_1'} +
sin\theta_{p_1}sin\theta_{p_1'}cos\phi_{q'}$ and \begin{eqnarray}
cos\theta_{p_1} &=& \frac {\frac {1} {2}q_1^2 + \frac {1} {4}q^2-{\tilde
p}^2 }
  {q_1\sqrt{{\tilde p}^2 + \frac {3} {4}q^2 - \frac {3} {4}q_1^2 } } \cr
cos\theta_{p_1~'} &=& \frac {p~'^2 - \frac {1} {2}q_1^2 - \frac {1} {4}q~'^2
}
  {q_1\sqrt{{p~'}^2 + \frac {3} {4}q~'^2 - \frac {3} {4}q_1^2 } } ~.
   \label{aa1.12a}
 \end{eqnarray}
In a similar way one gets for $X_2$
\begin{eqnarray}
 {\rm X}_{\rm 2}(\tilde p q \tilde \beta, p' q' \beta ')
   &=& \frac{{32\pi ^2 }}{{\tilde pp'qq'}}
\int\limits_{q_1^{\min } }^{q_1^{\max } }
 {dq_1 \frac{(-1)^{l'} {\delta _{LL'} }}{{2L + 1}}\sum\limits_\mu
 {Y_{\tilde l\lambda }^{*L\mu } (\theta _{\tilde p} ,
 \phi _{\tilde p}  = 0,\theta _q ,\phi _q  = 0)} }  \cr
  &\times& \int\limits_0^{2\pi } {d\phi _{q'} t_c^R (T,T';
  \sqrt {\tilde p^2  + \frac{3}{4}q^2  - \frac{3}{4}q_1^2 } ,
  \sqrt {p~'^2  + \frac{3}{4}q~'^2  - \frac{3}{4}q_1^2 } ,
  - \hat p_1  \circ \hat p_1 ~',E(q_1 ))}  \cr
  &\times& Y_{l'\lambda '}^{L\mu } (\theta _{p~'} ,\phi _{p~'}
   = \phi _{q '}  + \pi ,\theta _{q~'} ,\phi _{q~'} ) ~.
   \label{aa1.13}
 \end{eqnarray}
In (\ref{aa1.12}) and (\ref{aa1.13}) the limits of integrations
over $q_1$ are
\begin{eqnarray}
 q_1^{\min }  &=& \max \left\{ {\left| {\tilde p - \frac{q}{2}} \right|,
 \left| {p' - \frac{{q'}}{2}} \right|} \right\}
 \cr
 q_1^{\max }  &=& \min \left\{ {\tilde p + \frac{q}{2}
 ,p' + \frac{{q'}}{2}} \right\} ~.
 \label{aa1.14}
 \end{eqnarray}
The limits (\ref{aa1.14}) restrict the region of $p'$-values at given
$\tilde p$, $q$ and $q'$, for which $X_1$ and $X_2$ are nonzero.
For $q' \le 2|\tilde p - \frac {q} {2}|$ this requires that $p'
\in \left( {\left| {\tilde p - \frac{q}{2}} \right| -
\frac{{q'}}{2},\tilde p + \frac{q}{2} + \frac{{q'}}{2}} \right)$, 
 for $2|\tilde p - \frac {q} {2}| < q' \le 2(\tilde p + \frac {q}
{2})$:  $p' \in \left( 0,\tilde p + \frac{q}{2} +
\frac {q'}{2} \right)$, and for
 $q' > 2|\tilde p - \frac {q} {2}|$ and  $q' > 2(\tilde p + \frac {q}
{2})$:  $p' \in \left( \frac {q'} {2} - \left(
\tilde p + \frac {q} {2} \right ) , \frac {q'} {2} + \tilde p +
\frac{q}{2} \right)$. Inserting $X_1$ and $X_2$ from
(\ref{aa1.12}) and (\ref{aa1.13}) into (\ref{aa1.3}) and
(\ref{aa1.4}) and using the orthogonality of the CG-coefficients one
gets finally
\begin{eqnarray}
&&\left\langle {pq\alpha } \right|t_{N + c}^R PG_0 t_c^R PG_0
\left| {\alpha '} \right\rangle \left\langle {\alpha '}
\right|T\left| {\Phi } \right\rangle  = \sum\limits_{\tilde
\alpha } {\int {\tilde p^2 d\tilde p ~t_{\alpha \tilde \alpha }^{N
+ c,R} (p,\tilde p;E(q))G_0 (\tilde p,q)}} \cr && \sum\limits_{\tilde
\beta } {\left\langle {{\tilde \alpha }}
 \mathrel{\left | {\vphantom {{\tilde \alpha } {\tilde \beta }}}
 \right. \kern-\nulldelimiterspace}
 {{\tilde \beta }} \right\rangle \sum\limits_{\alpha ',\beta '}
 {\left\langle {{\beta '}}
 \mathrel{\left | {\vphantom {{\beta '} {\alpha '}}}
 \right. \kern-\nulldelimiterspace}
 {{\alpha '}} \right\rangle
\delta _{SS'} \delta _{\mu \mu '} \delta _{M_T , - 1/2} \delta
_{M_{T'} , - 1/2} 6\sqrt { {\hat t} {\hat t '}}
 \left\{
{\begin{array}{*{20}c}
   {1/2}  \\
   {1/2}  \\
\end{array}\begin{array}{*{20}c}
   {1/2}  \\
   T  \\
\end{array}\begin{array}{*{20}c}
   t  \\
   1  \\
\end{array}} \right\}\left\{ {\begin{array}{*{20}c}
   {1/2}  \\
   {1/2}  \\
\end{array}\begin{array}{*{20}c}
   {1/2}  \\
   {T'}  \\
\end{array}\begin{array}{*{20}c}
   1  \\
   {t'}  \\
\end{array}} \right\}}} \cr
&& \int {p'^2 dp'q'^2} dq' \left[ {( - 1)^{1 + s + t'} \sqrt {\hat s\hat s'}
\left\{
{\begin{array}{*{20}c}
   {1/2}  \\
   {1/2}  \\
\end{array}\begin{array}{*{20}c}
   {1/2}  \\
   S  \\
\end{array}\begin{array}{*{20}c}
   s  \\
   {s'}  \\
\end{array}} \right\}} X_1 + \delta _{ss'} X_2 \right]
 G_0
(p',q') <p'q'\alpha '|T| \Phi> ~.
  \label{aa1.15}
\end{eqnarray}
Performing the analogous steps and using
\begin{eqnarray}
 &&\left\langle {{{\rm p'q'}\alpha {\rm '}}}
 \mathrel{\left | {\vphantom {{{\rm p'q'}\alpha {\rm '}} \Phi }}
 \right. \kern-\nulldelimiterspace}
 {\Phi } \right\rangle {\rm  = }\frac{{\delta {\rm
 (q' - q}_{\rm 0} )}}{{{\rm q'}^{\rm 2} }}\delta _{{\rm t0}}
 \delta _{{\rm T1/2}} \delta _{{\rm j1}} \delta _{{\rm s1}}
 \varphi _l {\rm (p')}\sqrt {\frac{{2\lambda
  + 1}}{{4\pi }}} \left\langle {{1m_d Im_n }}
 \mathrel{\left | {\vphantom {{1m_d Im_n } {Jm_n  + m_d }}}
 \right. \kern-\nulldelimiterspace}
 {{Jm_n  + m_d }} \right\rangle \left\langle {{\lambda 0\frac{1}{2}m_n }}
 \mathrel{\left | {\vphantom {{\lambda 0\frac{1}{2}m_n } {Im_n }}}
 \right. \kern-\nulldelimiterspace}
 {{Im_n }} \right\rangle  \cr
 && \equiv \frac{{\delta {\rm (q' - q}_{\rm 0} )}}{{{\rm q'}^{\rm 2} }}
  \delta _{\alpha '\alpha _0 } \varphi _{\alpha _0 } (p')C_{\alpha _0 } ~,
  \label{aa1.16}
 \end{eqnarray}
 where it is assumed that the incoming nucleons momentum is
 parallel to the z-axis,
one gets the second matrix element $\left\langle {pq\alpha }
\right|t_{N+c}^RPG_0 t_c^R P \left| {\Phi } \right\rangle $:
\begin{eqnarray}
&&\left\langle {pq\alpha } \right|t_{N + c}^R PG_0 t_c^R P \left|
{\Phi } \right\rangle  = \sum\limits_{\tilde \alpha } {\int
{\tilde p^2 d\tilde p ~t_{\alpha \tilde \alpha }^{N + c,R}
(p,\tilde p;E(q))G_0 (\tilde p,q)}} \cr && \sum\limits_{\tilde \beta
} {\left\langle {{\tilde \alpha }}
 \mathrel{\left | {\vphantom {{\tilde \alpha } {\tilde \beta }}}
 \right. \kern-\nulldelimiterspace}
 {{\tilde \beta }} \right\rangle \sum\limits_{\alpha _0,\beta _0}
 {\left\langle {{\beta _0}}
 \mathrel{\left | {\vphantom {{\beta _0} {\alpha _0}}}
 \right. \kern-\nulldelimiterspace}
 {{\alpha _0}} \right\rangle C_{\alpha _0}
 \delta _{SS'} \delta _{\mu \mu '} \delta
_{M_T , - 1/2} \delta _{M_{T'} , - 1/2} 3\sqrt {\hat t\hat t'}
 \left\{ {\begin{array}{*{20}c}
   {1/2}  \\
   {1/2}  \\
\end{array}\begin{array}{*{20}c}
   {1/2}  \\
   T  \\
\end{array}\begin{array}{*{20}c}
   t  \\
   1  \\
\end{array}} \right\}}}
 \cr
&&\int p'^2 dp' \varphi_{\alpha_0}(p')
 \left[ (-)^{1 + s} \sqrt {\hat s\hat s'} \left\{
{\begin{array}{*{20}c}
   {1/2}  \\
   {1/2}  \\
\end{array}\begin{array}{*{20}c}
   {1/2}  \\
   S  \\
\end{array}\begin{array}{*{20}c}
   s  \\
   {s'}  \\
\end{array}} \right\} X_1(\tilde p q \tilde {\beta}, p'q_0 \beta_0)
+ \delta _{ss'} X_2(\tilde p q \tilde {\beta}, p'q_0 \beta_0)
 \right] ~,
  \label{aa1.17}
\end{eqnarray}
where $|\alpha_0>=|(l's')j'(\lambda'1/2)I'(j'I')J(t'1/2)T'=1/2>$, $l'=0,2$,
$t'=0$, $T'=\frac {1} {2}$,
$s'=j'=1$ and
\begin{eqnarray}
 {\rm C}_{\alpha _0 }  = \sqrt
{\frac{{2\lambda ' + 1}}{{4\pi }}} \left\langle {{1m_d I'm_n }}
 \mathrel{\left | {\vphantom {{1m_d I'm_n } {Jm_n  + m_d }}}
 \right. \kern-\nulldelimiterspace}
 {{Jm_n  + m_d }} \right\rangle \left\langle {{\lambda '0\frac{1}{2}m_n }}
 \mathrel{\left | {\vphantom {{\lambda '0\frac{1}{2}m_n } {I'm_n }}}
 \right. \kern-\nulldelimiterspace}
 {{I'm_n }} \right\rangle  ~.
 \label{aa1.18}
\end{eqnarray}

\section{Partial wave projected transition matrix elements: $<\vec p \vec q
|\alpha'><\alpha'|T|\Phi>$ and $<\vec p \vec q |\alpha'><\alpha'|PT|\Phi>$}
\label{b1}

Including spin ($m_i$) and isospin ($\nu_i$) projections of the
nucleons one gets for the kinematically complete breakup
configuration specified by Jacobi momenta ($\vec p$, $\vec q$) the
following contribution to the transition matrix element:
\begin{eqnarray}
&&\left\langle {\vec p\vec qm_1 m_2 m_3 \nu _1 \nu _2 \nu _3 }
\right|\sum\limits_\alpha  {\int {p'^2 dp'q'^2 dq'} \left|
{p'q'\alpha } \right\rangle } \left\langle {p'q'\alpha }
\right|T\left| {\Phi } \right\rangle  = \cr &&
\sum\limits_{J^\pi } {\sum\limits_{l\lambda L} {Y_{l\lambda }^{LM
- m_1  - m_2  - m_3 } (\hat p\hat q)\sum\limits_{jsIt}
{\sum\limits_S {\sqrt {\hat j\hat I\hat L\hat S} \left\{
{\begin{array}{*{20}c}
   l & s & j  \\
   \lambda  & {\frac{1}{2}} & I  \\
   L & S & J  \\
\end{array}} \right\}} } } } \cr
&& \left\langle {LSJ,M - m_1  - m_2  - m_3 ,m_1  + m_2  + m_3 }
\right\rangle \left\langle {s\frac{1}{2}S,m_2  + m_3 ,m_1 }
\right\rangle \cr &&\left\langle {\frac{1}{2}\frac{1}{2}s,m_2, m_3
} \right\rangle \left\langle {t\frac{1}{2}T,\nu _2  + \nu _3 ,\nu
_1 } \right\rangle \left\langle {\frac{1}{2}\frac{1}{2}t,\nu _2,
\nu _3 } \right\rangle \left\langle {pq\alpha } \right|T\left|
{\Phi } \right\rangle
\label{eq.a1.1}
\end{eqnarray}
where the incoming state $\left| {\Phi } \right\rangle \equiv
\left| {\vec q_0 ,\frac{1}{2}\mu } \right\rangle \left| {\varphi
_d ,1m_d } \right\rangle$ is composed from the relative
nucleon-deuteron motion with momentum $\vec q_0 \left\| z
\right.$, the deuteron wave function $\varphi _{d}$, and
$M=\mu+m_d$ is a sum of spin projections of the incoming nucleon
and deuteron.

The corresponding contribution to the elastic scattering
transition amplitude is:
\begin{eqnarray}
&&\left\langle {\Phi'} \right|P\sum\limits_{\alpha '} {\int
{p'^2 dp'q'^2 dq'\left| {p'q'\alpha '} \right\rangle } }
\left\langle {p'q'\alpha '} \right|T\left| {\Phi }
\right\rangle  = \cr &&\sum\limits_{J^\pi  M} {\sum\limits_{\alpha
_d 'l_0 \lambda _0 I_0 }  {\left\langle {1I_0 J,m_d ',M - m_d '}
\right\rangle \left\langle {\lambda _0 \frac{1}{2}I_0 ,M - m_d ' -
\mu ',\mu '} \right\rangle Y_{\lambda _0 ,M - m_d ' - \mu '} (\hat
q_0 ')}} \cr && {{ \int {q'^2 dq'\int\limits_{ - 1}^1 {dx\varphi
_{l_0 } (\pi _1 )\frac{{G_{\alpha _0 ,\alpha '} (q_0 q'x)}}{{\pi
_1^{l_0 } \pi _2^{l_{\alpha _d '} } }}\left\langle {\pi _2
q'\alpha '} \right|T\left| {\Phi } \right\rangle } } } }
\label{eq.a1.2}
\end{eqnarray}
where $|\alpha_d>=|(l_01)1(\lambda_0 \frac {1} {2})I_0
(1I_0)J,(\frac{1} {2} 0)T=\frac {1} {2} >$, $l_0 =0,2$,
$\varphi_{0,2}$ is the deuteron wave function, and $\pi _1  = \sqrt
{q'^2  + \frac{1}{4}q_0^2  + q'q_0 x}$,
 $\pi _2  = \sqrt {q_0 ^2  + \frac{1}{4}q'^2  + q'q_0 x}$.

\section{$\left\langle {\Phi' }
\right|PG_0^{-1}\left| {\Phi } \right\rangle$ and matrix elements with
3-dimensional Coulomb t-matrix: $\left\langle {\vec p\vec
q~ m_i \nu_i} \right|t_c^R P\left|
{\Phi } \right\rangle$ and $\left\langle {\Phi' } \right|Pt_c^R
P\left| {\Phi } \right\rangle$}
\label{c1}

Using
\begin{eqnarray}
P\left| {\vec p\vec qm_1 m_2 m_3 \nu _1 \nu _2 \nu _3 }
\right\rangle  &=& \left| { - \frac{1}{2}\vec p + \frac{3}{4}\vec
q, - \vec p - \frac{1}{2}\vec q,m_3 m_1 m_2 \nu _3 \nu _1 \nu _2 }
\right\rangle \cr
 &+& \left| { - \frac{1}{2}\vec p - \frac{3}{4}\vec
q,\vec p - \frac{1}{2}\vec q,m_2 m_3 m_1 \nu _2 \nu _3 \nu _1 }
\right\rangle
\label{eq.a2.1}
\end{eqnarray}
and
\begin{eqnarray}
&&\left\langle {\vec p\vec qm_1 m_2 m_3 \nu _1 \nu _2 \nu _3 }
\right|\left. {\Phi  } \right\rangle  \equiv \left\langle {\vec p\vec
qm_1 m_2 m_3 \nu _1 \nu _2 \nu _3 } \right|\left. {\varphi _d 1m_d
t = 00\frac{1}{2}m_N \frac{1}{2}m_t \vec q_0 } \right\rangle  =
\cr && \sum\limits_{L = 0,2} {\varphi _L (p)Y_{L,M = m_d  - m_2  -
m_3 } (\hat p)\left\langle {L11,M,m_d  - M} \right\rangle }\cr
&&{\left\langle {\frac{1}{2}\frac{1}{2}1,m_2 m_3 m_d  - M}
\right\rangle } \left\langle {\frac{1}{2}\frac{1}{2}0,\nu _2 \nu
_3 0} \right\rangle \delta _{m_1 m_N } \delta _{\nu _1 m_t }
\delta (\vec q - \vec q_0 ) \label{eq.a2.2}
\end{eqnarray}
one gets the following expression for the contribution of the
matrix element of the $t_c^RP$  to the breakup transition
amplitude
\begin{eqnarray}
&&\langle \vec p \vec q m_i \nu_i \vert t_c^RP \vert \Phi 
\rangle = \cr &&  \delta_{m_3,m_N}  \delta_{\nu_3,m_{t_N}}
\delta_{\nu_3,+\frac {1} {2} }  \delta_{\nu_2,+\frac {1} {2} }
 \sum_L{ \varphi_L(\vert \vec q + \frac {1} {2} \vec q_ 0 \vert )}
  Y_{L,m_d-m_1-m_2}(\hat {\vec q + \frac {1} {2} \vec q _0}~)\cr
 && (L11, m_d-m_1-m_2,m_1+m_2)  \cr
&& \times (\frac {1} {2}  \frac {1} {2} 1,m_1 m_2)
 \frac {\sqrt{2}}  {2} (-)^{ \frac {1} {2}-\nu_1} \delta_{\nu_1,-\nu_2}
 ~ \langle \vec p~ \vert t_c^R(E - \frac {3}{4m} q^2)
\vert -\frac {1} {2}\vec q - \vec q_0 \rangle + \cr &&
\delta_{m_2,m_N}  \delta_{\nu_2,m_{t_N}}
\delta_{\nu_2,+\frac {1} {2} }  \delta_{\nu_3,+\frac {1} {2} }
 \sum_L {\varphi_L(\vert \vec q + \frac {1} {2} \vec q_0 \vert )}
  Y_{L,m_d-m_3-m_1}(\hat{ {-\vec q - \frac {1} {2} \vec q_0}}~) \cr
&& (L11, m_d-m_3-m_1,m_3+m_1)  \cr && \times (\frac {1} {2} \frac
{1} {2} 1, m_3 m_1)
 \frac {\sqrt{2}}  {2} (-)^{ \frac {1} {2}-\nu_3} \delta_{\nu_3,-\nu_1}
 ~ \langle \vec p~ \vert t_c^R(E - \frac {3} {4m} q^2)
\vert \frac {1} {2} \vec q + \vec q_0 \rangle ~. \label{eq.a2.3}
\end{eqnarray}
The corresponding contribution to  the elastic scattering transition
amplitude from $Pt_c^RP$ is
\begin{eqnarray}
&&\langle \Phi' \vert Pt_c^RP \vert \Phi \rangle = \cr
&&\sum_{m_1 m_2} \sum_{L,\bar L} \int d\vec q \varphi_L(\vert \vec q +
{\frac {1} {2} }\vec q~'_0 \vert ) Y^*_{L,m_{d'}-m_1-m_2}(\hat {
 {\vec q + {\frac {1} {2}}\vec
q~'_0} }~) \varphi_{\bar L}(\vert \vec q + {\frac {1} {2}}\vec q_0 \vert )
Y_{\bar L,m_d-m_1-m_2}(\hat { {\vec q + {\frac {1} {2}}\vec q_0} }~) \cr
&&(L11\vert m_{d'}-m_1-m_2,m_1+m_2,m_d')
(\frac{1}{2} m_1 \frac {1}{2} m_2 \vert 1 m_1+m_2) \cr
&&(\bar L 1 1 \vert m_d-m_1-m_2,m_1+m_2,m_d)
(\frac{1}{2} m_1 \frac {1}{2} m_2 \vert 1 m_1+m_2)
\delta_{m_{N'},m_N} \delta_{m_{t_{N'}},m_{t_N}} \cr
&&t_c^R(\vert
{\frac {1} {2}}\vec q + \vec q~'_0 \vert, \vert {\frac {1} {2}}\vec q +
\vec q_0 \vert,
 cos\theta={\frac{  { {\frac {1} {4}}q^2 + \vec q_0 \cdot \vec q~'_0 +
 {\frac {1} {2}}\vec q \cdot (\vec q_0 + \vec q~'_0) }
} { { \vert {\frac {1} {2}}\vec q + \vec q~'_0 \vert \vert
{\frac {1} {2}}\vec q + \vec q_0 \vert } } }
; E-{\frac{3} {4m}}q^2) \cr &&
- \sum_{m_1} \sum_{L,\bar L} \int d\vec q \varphi_L(\vert \vec q +
{\frac {1} {2}}\vec q~'_0 \vert ) Y^*_{L,m_{d'}-m_1-m_N}(\hat {
 {\vec q + {\frac {1} {2}}\vec
q~'_0} }~) \varphi_{\bar L}(\vert \vec q + {\frac {1} {2}}\vec q_0 \vert )
Y_{\bar L,m_d-m_{N'}-m_1}(\hat { {\vec q + {\frac {1} {2}}\vec
q_0}}~)
\cr &&(L11 \vert m_{d'}-m_1-m_N,m_1+m_N,m_d')
(\frac{1}{2} m_1 \frac {1}{2} m_N \vert 1 m_1+m_N) \cr
&& (\bar L 1 1 \vert
m_d-m_{N'}-m_1, m_{N'}+m_1,m_d)
(\frac{1}{2} m_{N'} \frac {1}{2} m_1 \vert 1 m_{N'}+m_1)
\delta_{m_{t_{N}},m_{t_{N'}}} \cr
&&t_c^R(\vert {\frac {1} {2}}\vec q + \vec q~'_0
\vert, \vert {\frac {1} {2}}\vec q + \vec q_0 \vert,
 cos\theta= {\frac { { {\frac {1} {4}}q^2 + \vec q_0 \cdot \vec q~'_0 +
 {\frac {1} {2}}\vec q \cdot (\vec q_0 + \vec q~'_0) } }
 { { \vert {\frac {1} {2}}\vec q + \vec q~'_0 \vert \vert
{\frac {1} {2}}\vec q + \vec q_0 \vert  } } }; E-{\frac {3} {4m}}q^2)
\label{eq.a2.4}
\end{eqnarray}
where $\langle \vec p~ \vert t_c^R \vert \vec p~' \rangle \equiv
t_c^R(\vert \vec p~ \vert, \vert \vec p~' \vert, \hat p \cdot \hat
p~')$.

The matrix element $<\Phi'|PG_0^{-1}|\Phi>$ is given by
\begin{eqnarray}
&&\left\langle {\Phi'} \right|PG_0^{ - 1} \left| {\Phi }
\right\rangle  =  - \left[ {\frac{{q_0^2 }}{{m_p }}(\frac{5}{4} +
\cos \theta ) + \left| {E_d } \right|} \right]\sum\limits_{LL'm_2
} {\varphi _{L'} (\left| {\vec q_0  + \frac{1}{2}\vec q_0~ '}
\right|)}\cr &&{Y_{L',M' = m_d ' - m_N  - m_2 } (\hat{ \vec q_0  +
\frac{1}{2}\vec q_0~ '~}~)\left\langle {L'11,M',m_d ' - M'}
\right\rangle \left\langle {\frac{1}{2}\frac{1}{2}1,m_N m_2 m_d '
- M'} \right\rangle } \cr
&&\varphi _L (\left| {\frac{1}{2}\vec q_0
+ \vec q_0~ '} \right|)Y_{L,M = m_d  - m_2  - m_N }
(\hat {\frac{1}{2}\vec q_0  + \vec q_0~ '~ }~)\left\langle {L11,M,m_d  - M}
\right\rangle \left\langle {\frac{1}{2}\frac{1}{2}1,m_2 m_N m_d  -
M} \right\rangle ~.
 \label{eq.a2.5}
\end{eqnarray}

\section{Matrix elements with 3-dimensional Coulomb t-matrix:
$<\vec p \vec q~m_i \nu_i |t_c^RPG_0|\alpha'><\alpha'|T|\Phi>$
and $<\Phi'|Pt_c^RPG_0|\alpha'><\alpha'|T|\Phi>$}
\label{d1}

Inserting two complete set of states $|\vec p \vec q m_i \nu_i>$
($m_i$ and $\nu_i$ are nucleons spin- and isospin-projections) and
$|pq\alpha>$ one gets
\begin{eqnarray}
 \left\langle {\vec p\vec qm_i \nu _i } \right|t_c^R PG_0 T\left|
 \Phi  \right\rangle  &=& \int {d\vec p~'
t_c^R (\vec p,\vec p~';E(q))\sum\limits_\alpha
 {\int {dp''p''^2 } } } dq''q''^2 \left\langle {\vec p~'\vec qm_i \nu _i }
  \right|P\left| {p''q''\alpha } \right\rangle  \cr
 &\times& G_0 (p'',q'')\left\langle {p''q''\alpha } \right|T\left| \Phi
 \right\rangle ~.
\label{a.d.1}
 \end{eqnarray}

The permutation matrix element $\left\langle {\vec p~'\vec qm_i
\nu _i } \right|P\left| {p''q''\alpha } \right\rangle$ is given by
\begin{eqnarray}
&& \left\langle {\vec p~' \vec q m_i \nu _i } \right|P\left|
{p''q''\alpha } \right\rangle  =
 \sum\limits_{m_i ',\nu _i '} {\int {d\vec p_4 d\vec q_4 } }
\left\langle {\vec p~' \vec q m_i \nu _i }
  \right|P\left| {\vec p_4 \vec q_4 m_i '\nu _i '} \right\rangle
\left\langle
  {\vec p_4 \vec q_4 m_i '\nu _i '} \right| {p''q''\alpha } \left.
\right\rangle = \cr
 &&\sum\limits_{m_i ',\nu _i '} {\int {d\vec p_4 d\vec q_4 } } \left[ \delta
 \left( {\vec p~' - \vec {\pi_1}} \right)
\delta
 \left( {\vec p_4  + \vec {\pi_2}} \right)\delta _{m_1 m_3 '}
 \delta _{m_2 m_1 '} \delta _{m_3 m_2 '} \delta _{\nu _1 \nu _3 '}
 \delta _{\nu _2 \nu _1 '} \delta _{\nu _3 \nu _2 '}\right.  \cr
&&+ \delta \left( {\vec p~' + \vec {\pi_1}} \right)\delta
 \left( {\vec p_4   - \vec {\pi_2}} \right)
 \delta _{m_1 m_2 '} \delta _{m_2 m_3 '} \delta _{m_3 m_1 '}
 \delta _{\nu _1 \nu _2 '} \delta _{\nu _2 \nu _3 '}
\delta _{\nu _3 \nu _1 '} ]
 \cr
&& \sum\limits_{LS} {\sqrt {\hat j\hat I\hat L\hat S} } \left\{
{\begin{array}{*{20}c}
   l & s & j  \\
   \lambda  & {\frac{1}{2}} & I  \\
   L & S & J  \\
\end{array}} \right\}\sum\limits_{M_L } {\left\langle {{\frac{1}{2}m_2 '
\frac{1}{2}m_3 '}}
 \mathrel{\left | {\vphantom {{\frac{1}{2}m_2 '
\frac{1}{2}m_3 '} {sm_2 ' + m_3 '}}}
 \right. \kern-\nulldelimiterspace}
 {{sm_2 ' + m_3 '}} \right\rangle } \left\langle {{sm_2 ' + m_3 '
\frac{1}{2}m_1 '}}
 \mathrel{\left | {\vphantom {{sm_2 ' + m_3 '\frac{1}{2}m_1 '}
{Sm_1 ' + m_2 ' + m_3 '}}}
 \right. \kern-\nulldelimiterspace}
 {{Sm_1 ' + m_2 ' + m_3 '}} \right\rangle  \cr
&& \left\langle {{\frac{1}{2}\nu _2 '\frac{1}{2}\nu _3 '}}
 \mathrel{\left | {\vphantom {{\frac{1}{2}\nu _2 '\frac{1}{2}\nu _3 '}
{t\nu _2 ' + \nu _3 '}}}
 \right. \kern-\nulldelimiterspace}
 {{t\nu _2 ' + \nu _3 '}} \right\rangle
\left\langle {{t\nu _2 ' + \nu _3 '\frac{1}{2}\nu _1 '}}
 \mathrel{\left | {\vphantom {{t\nu _2 ' + \nu _3 '
\frac{1}{2}\nu _1 '} {T\nu _1 ' + \nu _2 ' + \nu _3 '}}}
 \right. \kern-\nulldelimiterspace}
 {{T\nu _1 ' + \nu _2 ' + \nu _3 '}}
\right\rangle \left\langle {{LM_L Sm_1 ' + m_2 ' + m_3 '}}
 \mathrel{\left | {\vphantom {{LM_L Sm_1 ' + m_2 ' + m_3 '} {Jm_p  + m_d }}}
 \right. \kern-\nulldelimiterspace}
 {{Jm_p  + m_d }} \right\rangle  \cr
&& \frac{{\delta \left( {p'' - p_4 } \right)}}{{p''^2
}}\frac{{\delta \left( {q'' - q_4 } \right)}}{{q''^2 }}Y_{l\lambda
}^{LM_L } \left( {\hat p_4,\hat q_4} \right)
 \label{a.d.2}
 \end{eqnarray}
where
\begin{eqnarray}
\vec {\pi_1} &=& \frac {1} {2}\vec q + \vec q_4 \cr
\vec {\pi_2}
&=& \vec q + \frac {1} {2}\vec q_4 ~.
 \label{a.d.3}
 \end{eqnarray}

Performing in (\ref{a.d.2}) integrations over $p''$ and  $q''$ leads to
\begin{eqnarray}
 &&\left\langle {\vec p\vec qm_1 m_2 m_3 \nu _1 \nu _2 \nu _3 }
 \right|t_c^R PG_0 T\left| \Phi  \right\rangle
 = \int d\vec p~' d\vec p_4 d\vec q_4
 t_c^R (\vec p,\vec p~';E(q)) \cr
 && \times \sum\limits_{\alpha m_i' \nu_i'}
\left[ \delta
 \left( {\vec p~' - \vec {\pi_1}} \right)
\delta
 \left( {\vec p_4  + \vec {\pi_2}} \right)\delta _{m_1 m_3 '}
 \delta _{m_2 m_1 '} \delta _{m_3 m_2 '} \delta _{\nu _1 \nu _3 '}
 \delta _{\nu _2 \nu _1 '} \delta _{\nu _3 \nu _2 '}\right.  \cr
&&+ \delta \left( {\vec p~' + \vec {\pi_1}} \right)\delta
 \left( {\vec p_4   - \vec {\pi_2}} \right)
 \delta _{m_1 m_2 '} \delta _{m_2 m_3 '} \delta _{m_3 m_1 '}
 \delta _{\nu _1 \nu _2 '} \delta _{\nu _2 \nu _3 '}
\delta _{\nu _3 \nu _1 '} ]
 \cr
&& \sum\limits_{LS} {\sqrt {\hat j\hat I\hat L\hat S} } \left\{
{\begin{array}{*{20}c}
   l & s & j  \\
   \lambda  & {\frac{1}{2}} & I  \\
   L & S & J  \\
\end{array}} \right\}\sum\limits_{M_L }
{\left\langle {{\frac{1}{2}m_2 '\frac{1}{2}m_3 '}}
 \mathrel{\left | {\vphantom {{\frac{1}{2}m_2 '
\frac{1}{2}m_3 '} {sm_2 ' + m_3 '}}}
 \right. \kern-\nulldelimiterspace}
 {{sm_2 ' + m_3 '}} \right\rangle }
\left\langle {{sm_2 ' + m_3 '\frac{1}{2}m_1 '}}
 \mathrel{\left | {\vphantom {{sm_2 ' + m_3 '
\frac{1}{2}m_1 '} {Sm_1 ' + m_2 ' + m_3 '}}}
 \right. \kern-\nulldelimiterspace}
 {{Sm_1 ' + m_2 ' + m_3 '}} \right\rangle  \cr
&& \left\langle {{\frac{1}{2}\nu _2 '\frac{1}{2}\nu _3 '}}
 \mathrel{\left | {\vphantom {{\frac{1}{2}\nu _2 '
\frac{1}{2}\nu _3 '} {t\nu _2 ' + \nu _3 '}}}
 \right. \kern-\nulldelimiterspace}
 {{t\nu _2 ' + \nu _3 '}} \right\rangle
\left\langle {{t\nu _2 ' + \nu _3 '\frac{1}{2}\nu _1 '}}
 \mathrel{\left | {\vphantom {{t\nu _2 ' + \nu _3 '
\frac{1}{2}\nu _1 '} {T\nu _1 ' + \nu _2 ' + \nu _3 '}}}
 \right. \kern-\nulldelimiterspace}
 {{T\nu _1 ' + \nu _2 ' + \nu _3 '}}
\right\rangle \left\langle {{LM_L Sm_1 ' + m_2 ' + m_3 '}}
 \mathrel{\left | {\vphantom {{LM_L Sm_1 ' + m_2 ' + m_3 '} {JM_J }}}
 \right. \kern-\nulldelimiterspace}
 {{JM_J }} \right\rangle  \cr
&& Y_{l\lambda }^{LM_L } \left( {\hat p_4,\hat q_4} \right)
 G_0(p_4,q_4)
  \left\langle {p_4,q_4,\alpha } \right|T\left| \Phi  \right\rangle ~.
  \label{a.d.4}
 \end{eqnarray}

For channels $\alpha \ne \alpha_d$ the only singularity is the
$G_0$ singularity.  When $\vec p_4$ and $\vec q_4$ are taken as an
integration variables in (\ref{a.d.4}) $G_0$ becomes
\begin{eqnarray}
G_0(p_4,q_4)&=& G_0(p',q) = \frac {1} {E + i\epsilon - \frac {1}
{m}(p~'^2 + \frac {3} {4}q^2 )  }
 \label{a.d.5}
 \end{eqnarray}
and  the contribution to the matrix element $\left\langle {\vec
p\vec qm_1 m_2 m_3 \nu _1 \nu _2 \nu _3 }
 \right|t_c^R PG_0 T\left| \Phi  \right\rangle$ coming from
these channels is given by
\begin{eqnarray}
 &&\left\langle {\vec p\vec qm_1 m_2 m_3 \nu _1 \nu _2 \nu _3 }
 \right|t_c^R PG_0 T\left| \Phi  \right\rangle |_{\alpha \ne \alpha_d}
 = \delta_{\nu_2,+\frac {1}{2}} \delta_{\nu_3,+\frac {1}{2}}
 \int {p~'^2 dp~'\frac{1}{{E + i\varepsilon  - \frac{{p~'^2 }}{m} -
 \frac{{3q^2 }}{{4m}}}} 
 \int {d\phi _{p~'} d(\cos \theta _{p~'} )}} \cr  
&& t_c^R (\vec p,\vec p~';E(q)) 
 \sum\limits_{\alpha \ne \alpha_d
LS}
 {\sqrt {\hat j\hat I\hat L\hat S} \left\{ {\begin{array}{*{20}c}
   l & s & j  \\
   \lambda  & {\frac{1}{2}} & I  \\
   L & S & J  \\
\end{array}} \right\}\left\langle {{LM_L Sm_1  + m_2  + m_3 }}
 \mathrel{\left | {\vphantom {{LM_L Sm_1  + m_2  + m_3 } {Jm_p  + m_d }}}
 \right. \kern-\nulldelimiterspace}
 {{Jm_p  + m_d }} \right\rangle }  \cr
&& [\left\langle {{\frac{1}{2}m_3 \frac{1}{2}m_1 }}
 \mathrel{\left | {\vphantom {{\frac{1}{2}m_3
\frac{1}{2}m_1 } {sm_3  + m_1 }}}
 \right. \kern-\nulldelimiterspace}
 {{sm_3  + m_1 }} \right\rangle \left\langle {{sm_3  + m_1 \frac{1}{2}m_2 }}
 \mathrel{\left | {\vphantom {{sm_3  + m_1 \frac{1}{2}m_2 }
{Sm_1  + m_2  + m_3 }}}
 \right. \kern-\nulldelimiterspace}
 {{Sm_1  + m_2  + m_3 }} \right\rangle  \cr
&& \left\langle {{\frac{1}{2}\nu _3 \frac{1}{2}\nu _1 }}
 \mathrel{\left | {\vphantom {{\frac{1}{2}\nu _3
\frac{1}{2}\nu _1 } {t\nu _3  + \nu _1 }}}
 \right. \kern-\nulldelimiterspace}
 {{t\nu _3  + \nu _1 }} \right\rangle
\left\langle {{t\nu _3  + \nu _1 \frac{1}{2}\nu _2 }}
 \mathrel{\left | {\vphantom {{t\nu _3  + \nu _1 \frac{1}{2}\nu _2 }
 {T\nu _1  + \nu _2  + \nu _3 }}}
 \right. \kern-\nulldelimiterspace}
 {{T\nu _1  + \nu _2  + \nu _3 }} \right\rangle  \cr
&& \times Y_{l\lambda }^{LM_L } \left( { \hat {- \frac{1}{2}\vec p~' -
\frac{3}{4}\vec q}~~ ,\hat {\vec p~'
  - \frac{1}{2}\vec q}~~ } \right)\left\langle {| -
\frac{1}{2}\vec p~' - \frac{3}{4}\vec q~|,
  |\vec p~' - \frac{1}{2}\vec q~|,\alpha }
\right|T\left| \Phi  \right\rangle \cr
&&  + \left\langle {{\frac{1}{2}m_1 \frac{1}{2}m_2 }}
 \mathrel{\left | {\vphantom {{\frac{1}{2}m_1
\frac{1}{2}m_2 } {sm_1  + m_2 }}}
 \right. \kern-\nulldelimiterspace}
 {{sm_1  + m_2 }} \right\rangle \left\langle {{sm_1  + m_2 \frac{1}{2}m_3 }}
 \mathrel{\left | {\vphantom {{sm_1  + m_2 \frac{1}{2}m_3 }
{Sm_1  + m_2  + m_3 }}}
 \right. \kern-\nulldelimiterspace}
 {{Sm_1  + m_2  + m_3 }} \right\rangle  \cr
&& \left\langle {{\frac{1}{2}\nu _1 \frac{1}{2}\nu _2 }}
 \mathrel{\left | {\vphantom {{\frac{1}{2}\nu _1
\frac{1}{2}\nu _2 } {t\nu _1  + \nu _2 }}}
 \right. \kern-\nulldelimiterspace}
 {{t\nu _1  + \nu _2 }} \right\rangle
\left\langle {{t\nu _1  + \nu _2 \frac{1}{2}\nu _3 }}
 \mathrel{\left | {\vphantom {{t\nu _1  + \nu _2
\frac{1}{2}\nu _3 } {T\nu _1  + \nu _2  + \nu _3 }}}
 \right. \kern-\nulldelimiterspace}
 {{T\nu _1  + \nu _2  + \nu _3 }} \right\rangle  \cr
&&  \times Y_{l\lambda }^{LM_L } \left( { \hat {- \frac{1}{2}\vec p~'
  + \frac{3}{4}\vec q}~~, \hat {- \vec p~' - \frac{1}{2}\vec q}~~ } \right)
  \left\langle {| - \frac{1}{2}\vec p~' + \frac{3}{4}\vec q~|,| - \vec p~'
  - \frac{1}{2}\vec q~|,\alpha } \right|T\left| \Phi  \right\rangle ] ~,
\label{a.d.6}
 \end{eqnarray}
where $m_p$ and $m_d$ are spin projections of the incoming proton
and deuteron, and $M_L=m_p+m_d-m_1-m_2-m_3$. Integration over $p'$
can be performed numerically taking care of the pole in $G_0$ by
e.g. the subtraction method.

For $\alpha_d$ channels one can decompose
$G_0(p_4,q_4)\left\langle p_4 q_4 \alpha_d |T| \Phi \right \rangle$ in
(\ref{a.d.4}) in
the following way
\begin{eqnarray}
&&G_0(p_4,q_4)\left\langle p_4 q_4 \alpha_d |T| \Phi \right \rangle=
\frac {1} {\frac {3} {4m}(q_{max}^2-q_4^2) -\frac {1} {m}p_4^2 +
i\epsilon   }   \frac {\left\langle p_4 q_4 \alpha_d |\hat T| \Phi
\right \rangle} {\frac {3} {4m} (q_0^2-q_4^2) +i\epsilon} \cr
&&\frac {1} {\frac {3} {4m}(q_{max}^2-q_4^2) -\frac {1} {m}p_4^2 +
i\epsilon   }   \frac {\left\langle p_4 q_4 \alpha_d |\hat T| \Phi
\right \rangle} { |E_d| + \frac {1} {m} p_4^2} - \frac {1} {\frac
{3} {4m}(q_0^2-q_4^2)  + i\epsilon   } \frac {\left\langle p_4 q_4
\alpha_d |\hat T| \Phi \right \rangle} { |E_d| + \frac {1} {m} p_4^2} ~.
 \label{a.d.7}
 \end{eqnarray}
The first part in (\ref{a.d.7}) can be integrated in the same way as the
contribution from channels $\alpha \ne \alpha_d$, resulting in
expression (\ref{a.d.6}) with $\left\langle p q \alpha |T| \Phi \right
\rangle$ replaced by $\frac {\left\langle p q \alpha_d |\hat T| \Phi
\right \rangle} { |E_d| + \frac {1} {m} p^2}$.

Calculating  contribution from the second term in (\ref{a.d.7}) one
takes in  (\ref{a.d.3})
 $\vec p~'$ and $\vec p_4$ as an integration variables. This gives
the following result for  that term
\begin{eqnarray}
 &&\left\langle {\vec p\vec qm_1 m_2 m_3 \nu _1 \nu _2 \nu _3 }
 \right|t_c^R PG_0 T\left| \Phi  \right\rangle |_{\alpha_d}^{second}
 = - \delta_{\nu_2,+\frac {1}{2}} \delta_{\nu_3,+\frac {1}{2}}
\int {q_4^2 dq_4\frac{1}{{ \frac {3} {4m} (q_0^2-q_4^2) + i\varepsilon}}
 \int {d\phi _{q_4} d(\cos \theta _{q_4} )} } \cr
 && \sum\limits_{\alpha_d LS}
 {\sqrt {\hat j\hat I\hat L\hat S} \left\{ {\begin{array}{*{20}c}
   l & s & j  \\
   \lambda  & {\frac{1}{2}} & I  \\
   L & S & J  \\
\end{array}} \right\}\left\langle {{LM_L Sm_1  + m_2  + m_3 }}
 \mathrel{\left | {\vphantom {{LM_L Sm_1  + m_2  + m_3 } {Jm_p  + m_d }}}
 \right. \kern-\nulldelimiterspace}
 {{Jm_p  + m_d }} \right\rangle }  \cr
&& [\left\langle {{\frac{1}{2}m_3 \frac{1}{2}m_1 }}
 \mathrel{\left | {\vphantom {{\frac{1}{2}m_3
\frac{1}{2}m_1 } {sm_3  + m_1 }}}
 \right. \kern-\nulldelimiterspace}
 {{sm_3  + m_1 }} \right\rangle \left\langle {{sm_3  + m_1 \frac{1}{2}m_2 }}
 \mathrel{\left | {\vphantom {{sm_3  + m_1 \frac{1}{2}m_2 }
{Sm_1  + m_2  + m_3 }}}
 \right. \kern-\nulldelimiterspace}
 {{Sm_1  + m_2  + m_3 }} \right\rangle  \cr
&& \left\langle {{\frac{1}{2}\nu _3 \frac{1}{2}\nu _1 }}
 \mathrel{\left | {\vphantom {{\frac{1}{2}\nu _3
\frac{1}{2}\nu _1 } {t\nu _3  + \nu _1 }}}
 \right. \kern-\nulldelimiterspace}
 {{t\nu _3  + \nu _1 }} \right\rangle
\left\langle {{t\nu _3  + \nu _1 \frac{1}{2}\nu _2 }}
 \mathrel{\left | {\vphantom {{t\nu _3  + \nu _1 \frac{1}{2}\nu _2 }
 {T\nu _1  + \nu _2  + \nu _3 }}}
 \right. \kern-\nulldelimiterspace}
 {{T\nu _1  + \nu _2  + \nu _3 }} \right\rangle t_c^R (\vec
p,\frac {1} {2} \vec q + \vec q_4;E(q)) \cr
&& \times Y_{l\lambda }^{LM_L } \left( { \hat {-\vec q-\frac{1}{2}\vec
q_4}~~,
\hat q_4}
 \right)  \frac
{ \left\langle | -\vec q - \frac {1} {2}\vec q_4~|,q_4,\alpha_d
\right| \hat T \left| \Phi  \right\rangle }
{ |E_d| +\frac {1} {m} { \left| \vec q +
\frac {1} {2} \vec q_4 \right| }^2 } \cr
&&  + \left\langle {{\frac{1}{2}m_1 \frac{1}{2}m_2 }}
 \mathrel{\left | {\vphantom {{\frac{1}{2}m_1
\frac{1}{2}m_2 } {sm_1  + m_2 }}}
 \right. \kern-\nulldelimiterspace}
 {{sm_1  + m_2 }} \right\rangle \left\langle {{sm_1  + m_2 \frac{1}{2}m_3 }}
 \mathrel{\left | {\vphantom {{sm_1  + m_2 \frac{1}{2}m_3 }
{Sm_1  + m_2  + m_3 }}}
 \right. \kern-\nulldelimiterspace}
 {{Sm_1  + m_2  + m_3 }} \right\rangle  \cr
&& \left\langle {{\frac{1}{2}\nu _1 \frac{1}{2}\nu _2 }}
 \mathrel{\left | {\vphantom {{\frac{1}{2}\nu _1
\frac{1}{2}\nu _2 } {t\nu _1  + \nu _2 }}}
 \right. \kern-\nulldelimiterspace}
 {{t\nu _1  + \nu _2 }} \right\rangle
\left\langle {{t\nu _1  + \nu _2 \frac{1}{2}\nu _3 }}
 \mathrel{\left | {\vphantom {{t\nu _1  + \nu _2
\frac{1}{2}\nu _3 } {T\nu _1  + \nu _2  + \nu _3 }}}
 \right. \kern-\nulldelimiterspace}
 {{T\nu _1  + \nu _2  + \nu _3 }} \right\rangle  t_c^R (\vec
p,-\frac {1} {2} \vec q - \vec q_4 ;E(q))\cr
 &&  \times Y_{l\lambda }^{LM_L } \left( { \hat { \vec q
  + \frac{1}{2}\vec q_4 }~~ , \hat q_4 } \right)
  \frac { \left\langle {| \vec q + \frac{1}{2}\vec q_4 |,q_4,\alpha_d }
\right| \hat T \left| \Phi
  \right\rangle} {|E_d| +
\frac {1} {m}|\vec q + \frac {1} {2} \vec q_4 |^2  } ] ~.
\label{a.d.8}
 \end{eqnarray}
Again one has a simple pole which can be treated using e.g. the
 subtraction method.

For the exclusive breakup the calculation of the contributions (\ref{a.d.6})
and
(\ref{a.d.8}) must be performed for each complete geometry in a coordinate
system used when calculating $<pq\alpha|T|\Phi>$ where the
 z-  axis was taken to be parallel to the incoming proton momentum.

For the elastic scattering an additional integration has to be done
providing the following contribution to the elastic scattering amplitude
from the $<\Phi'|Pt_c^RPG_0|\alpha'><\alpha'|T|\Phi>$ term: 
\begin{eqnarray}
&&<\Phi'|Pt_c^RPG_0|\alpha'><\alpha'|T|\Phi> =
 [ - \int d\vec q \sum\limits_{m_1 m_2} \sum\limits_{LM}
\varphi _L (\left| {\vec q + \frac{1}{2}\vec q_0 ~'} \right|)
Y_{L,M }^* (\hat {\vec q + \frac{1}{2} \vec q_0~ '~ }~) \cr
&& \left\langle {L11,M,m_d'  - M,m_d'}
\right\rangle \left\langle {\frac{1}{2}\frac{1}{2}1,m_1 m_2 m_d'  -M}
 \right\rangle \frac {\sqrt{2}} {2}    \cr
&&  \left\langle {-\frac {1}{2}\vec q - \vec q_0~',   \vec q m_1 m_2 m_{N'},
 \nu _1=-\frac {1}{2} \nu _2=+ \frac {1}{2} ~ m_{t_{N'}}=+ \frac {1}{2} }
 \right|t_c^R PG_0 T\left| \Phi  \right\rangle \cr
&& + \sum\limits_{m_1 m_3} \sum\limits_{LM}
\varphi _L (\left| {-\vec q - \frac{1}{2}\vec q_0 ~'} \right|)
Y_{L,M }^* (\hat {-\vec q - \frac{1}{2} \vec q_0~ '~ }~) \cr
&& \left\langle {L11,M,m_d'  - M,m_d'}
\right\rangle \left\langle {\frac{1}{2}\frac{1}{2}1,m_3 m_1 m_d'  -M}
 \right\rangle  \frac {\sqrt{2}} {2}  \cr
&&  \left\langle {\frac {1}{2}\vec q + \vec q_0~',   \vec q m_1  m_{N'} m_3,
 \nu _1= -\frac {1}{2}  m_{t_{N'}}= + \frac {1}{2}\nu_3=+ \frac {1}{2} }
 \right|t_c^R PG_0 T\left| \Phi  \right\rangle  ~.
\label{a.d.9}
 \end{eqnarray}

\end{document}